\documentclass[preprint]{aastex}

\begin{document}

\title{Asymmetries in SN 2014J near maximum light revealed through spectropolarimetry}

\author{Amber L. Porter\altaffilmark{1,2}, Mark D. Leising\altaffilmark{2}, G. Grant Williams\altaffilmark{3,4}, Peter Milne\altaffilmark{3}, Paul Smith\altaffilmark{3}, Nathan Smith\altaffilmark{3}, Christopher Bilinski\altaffilmark{3}, Jennifer L. Hoffman\altaffilmark{5}, Leah Huk\altaffilmark{5}, Douglas C. Leonard\altaffilmark{6} }

\altaffiltext{1}{alporte@g.clemson.edu}
\altaffiltext{2}{Department of Physics and Astronomy, Clemson University, 118 Kinard Laboratory, Clemson, SC 29634.}
\altaffiltext{3}{Steward Observatory, University of Arizona, 933 N. Cherry Ave, Tucson, AZ 85721.}
\altaffiltext{4}{MMT Observatory, P.O. Box 210065, University of Arizona, Tucson, AZ 85721-0065.}
\altaffiltext{5}{Department of Physics and Astronomy, University of Denver, 2112 East Wesley Avenue, Denver, CO 80208.}
\altaffiltext{6}{Department of Astronomy, San Diego State University, PA-210, 5500 Campanile Drive, San Diego, CA 92182-1221.}

\begin{abstract}

We present spectropolarimetric observations of the nearby Type Ia SN 2014J in M82 over six epochs: +0, +7, +23, +51, +77, +109, and +111 days with respect to B-band maximum.  The strong continuum polarization, which is constant with time, shows a wavelength dependence unlike that produced by linear dichroism in Milky Way dust.  The observed polarization may be due entirely to interstellar dust or include a circumstellar scattering component.  We find that the polarization angle aligns with the magnetic field of the host galaxy, arguing for an interstellar origin.  Additionally, we confirm a peak in polarization at short wavelengths that would imply $R_V < 2 $ along the light of sight, in agreement with earlier polarization measurements.  For illustrative purposes, we include a two component fit to the continuum polarization of our +51 day epoch that combines a circumstellar scattering component with interstellar dust where scattering can account for over half of the polarization at $4000$ \AA.  Upon removal of the interstellar polarization signal, SN 2014J exhibits very low levels of continuum polarization.  Asymmetries in the distribution of elements within the ejecta are visible through moderate levels of time-variable polarization in accordance with the Si II 6355  \AA ~absorption line.  At maximum light, the line polarization reaches $\sim0.6$\% and decreases to $\sim0.4\%$ one week later.  This feature also forms a loop on the $q_{RSP}$-$u_{RSP}$ plane illustrating that the ion does not have an axisymmetric distribution.  The observed polarization properties suggest the explosion geometry of SN 2014J is generally spheroidal with a clumpy distribution of silicon.

\end{abstract}



\section{Introduction}

The ability to use Type Ia supernovae (SNe Ia) peak magnitudes as standardizable candles has made these explosions instrumental in measuring cosmic distances \citep{Riess,Perlmutter}.  A general agreement exists that SNe Ia are the thermonuclear explosion of white dwarf stars that explode in a binary system \citep{Hoyle}.  However, determining the true progenitor system, specifically, how to grow a white dwarf to the Chandrasekhar mass prior to exploding has been a challenge \citep[see][for recent reviews]{Howell_review,Maoz_ARAA}.  One of the commonly accepted scenarios is the single degenerate (SD) system where the white dwarf accretes matter from a non-degenerate companion \citep{Whelan}.  An alternative theory is the double degenerate (DD) channel in which a more massive white dwarf tidally disrupts its less massive companion and then accretes the material \citep{Iben,Webbink}.  SD explosion models can reasonably match observed light curves and spectra of SNe Ia \citep{Stehle_2002bo}, but we have yet to find conclusive proof of the companion star of these progenitor systems in pre- and post-explosion images \citep{Li_2011feProg, Schaefer}.  Meanwhile, the nearby explosion of SN 2011fe has lent support to the DD case as deep X-ray \citep{Margutti_2011fe} and radio \citep{Chomiuk_2011fe} observations indicate a clean local environment.  However, the significant asymmetry expected in violent-merger DD explosions \citep{Bulla} is not consistent with the modest levels of polarization measured in SN 2011fe \citep{SP2011fe}.  Thus, asymmetries present in the ejecta of extragalactic supernovae, which can be explored through spectropolarimetry, illuminate details about the explosion such as the overall asphericity and the stratification of elements in the outer ejecta layers \citep{SPReview}.  This can then be used to distinguish between the various explosion models and progenitor systems.

The supernova's light becomes linearly polarized when scattered by electrons which are abundant in the atmosphere.  A spherically symmetric distribution of electrons in a source that is not resolved causes the electric vectors to cancel completely and leads to zero net polarization \citep{Hoflich}.  Therefore, a non-null polarization measurement demands some level of asymmetry as it is the result of the unequal canceling of perpendicular electric field vectors.  Possible causes of asymmetry in SNe Ia are the encounter between the supernova ejecta and the evolved companion star in the SD progenitor scenario \citep{Marietta,Kasen_hole} or the disruption of a white dwarf in the DD case \citep{Livio,Bulla}.  Polarization may also be produced by irregularities in the ionization structure caused by clumps of the radioactive $^{56}$Ni in the ejecta \citep{Chugai}.

The continuum polarization of normal SNe Ia is typically measured to be quite low, $P \le 0.3\%$, suggesting a global geometry that is nearly spherical \citep{SPReview}.  Line polarization, on the other hand, indicates that the distribution of those specific ions is not uniform along the line of sight above the photosphere.  The Si II 6355 \AA ~and the Ca II near-IR triplet lines were significantly polarized in all well-studied SNe Ia with spectropolarimetric data to date near maximum light \citep{SP2001el,Leonard,SP2004S,SP2006X,SP2008fp}.  Some explosions also showed signs of line polarization across Si II 5051 \AA, Mg II 4471 \AA ~\citep{SP2006X}, S II near 5300 \AA ~\citep{SP2011fe}, and Fe II near 4800 \AA ~\citep{Leonard,SP2004S}, although all of these lines have shown less significance.  Therefore, we see that a number of intermediate-mass elements in SNe Ia have a clumpy distribution in the ejecta.  Both the continuum and line polarization show a temporal evolution which decreases toward a null level of polarization between maximum light and two weeks later indicating that SNe Ia become more spherically symmetric with time \citep{SPReview}.  The Si II 6355 \AA~ line polarization has also been used to show that brighter SNe Ia \citep{Wang_Sci} and those with lower expansion velocities \citep{Maund_uni} tend to be less polarized.

In addition to learning about the supernova itself, for cases where there is appreciable extinction, the polarized continua of SNe may enable a study of the host galaxies' dust properties as the wavelength dependence of the polarization has been related to the extinction law \citep{Serkowski1975}.  This method depends on the intrinsic polarization of the supernova being negligible compared to interstellar polarization (ISP) when the color excess $E(B-V)> 0.2~\mathrm{mag}$.  The wavelength-dependent extinction of the Milky Way is often characterized by $R_V$, where $R_V=A_V / E(B-V)$, which relates the selective extinction, $R_V$, to the total extinction, $A_V$, and the amount of reddening, $E(B-V)$.

Although it varies with line of sight, an $R_V \sim 3.1$ is often quoted to describe the average extinction law of the Milky Way \citep{CCM}.  A lower $R_V$ value leads to an extinction law that rises more steeply in the blue as compared to the Galaxy and is thought to be the result of an increased number of grains with radii $a < 0.1 \, \mathrm{\mu m}$ \citep{Kim}.  Several studies of well-observed SNe Ia have shown such steep extinction laws with the most extreme cases displaying an $R_V < 2$ \citep{XWang_2006X,Kris_2006,Elias-Rosa_2008,Elias-Rosa_2006}.  As this value is below the canonical value of the Milky Way, the dust obscuring these explosions is likely different from the interstellar medium of the Galaxy.  Spectropolarimetry of some highly reddened SNe Ia agree with this result as the wavelength dependence of their polarization implies $R_V < 2$ along their line of sight.  We discuss this in  more detail in Sec. \ref{Serkowski}.  These are important results when using SNe Ia as standardizeable candles in cosmology because \citet{Hicken} has shown that an $R_V = 3.1$ overestimates host galaxy extinction of the CfA mid-$z$ sample.  An $R_V = 1.7$ however, reduced the Hubble residuals for this sample.

Here we present optical spectropolarimetry of the normal SN Ia SN 2014J for 111 days starting at maximum light obtained with the SPOL Spectropolarimeter at Steward Observatory 2.3-m Bok and 6.5-m MMT telescopes.  The high level of continuum polarization allows us to analyze the dust properties of the host galaxy and the multi-epoch observations add to the growing number of well-studied SNe Ia for which the evolution of the polarization properties has been investigated in detail.

\subsection{SN 2014J}   \label{PreviousObs}
SN 2014J was discovered by \cite{Fossey} on 2014 January 21.805 (UT) in the nearby starburst galaxy M82.  The distance to the galaxy as determined by \citet{Dalcanton} is $D=3.5 \pm 0.3$ Mpc making SN 2014J the closest SN Ia in several decades.  Early Palomar Transient Factory spectra showed SN 2014J followed a similar evolution to supernovae such as SN 2011fe, but with slightly higher ejecta velocities \citep{Goobar}.  The authors noted, however, that strong attenuation at bluer wavelengths indicated that the supernova was suffering from a significant amount of extinction along the line of sight.  The maximum light Si II 6355 \AA ~velocity, measured from the middle of the absorption line, was $v_{SiII,0} = -12,000\,\mathrm{km\,s^{-1}}$ \citep{Marion} placing SN 2014J slightly inside the \citet{Wang_golden} high velocity (HV) classification (the HV group is defined as having $v_{SiII,0} < -11,800\,\mathrm{km\,s^{-1}}$) which is a proxy for the velocity gradient when spectroscopic sampling is sparse.  However, using their spectra from [-0.4,+9.1] days, \citet{Marion} found a velocity gradient $\dot v=42 \,\mathrm{km\,s^{-1}\,day^{-1}}$, placing SN 2014J in the low velocity gradient (LVG) group of \citet{Benetti}.  \citet{Foley} measured a peak brightness $B_{max}=11.85 \pm 0.02$ mag on $JD=2456690.5 \pm 0.2$ and a decline rate parameter $\Delta m_{15}(B)_{\mathrm{obs}}=0.95 \pm 0.01 \, \mathrm{mag}$ (both values uncorrected for reddening).

\subsection{Extinction law of SN 2014J}
The intrinsic colors of SNe Ia are believed to be nearly homogeneous with a slight dependence on light curve shape \citep{Phillips_Color}.  Based upon this assumption, the extinction law of a supernova is generally determined by comparing the observed colors or spectra to unreddened standards.  The low $R_V$ value exhibited by highly extincted SNe Ia would imply that the interstellar dust along the line of sight has a small average grain size if the extinction is dominated by ISM dust \citep{Whittet_book}.  However, \citet{WangLE} showed that these low values can possibly be explained by dust in a circumstellar environment scattering and reddening the supernova light.  Simulations by \citet{Goobar_Scat} later showed scattering and reddening could give rise to a power-law wavelength-dependent extinction law of the form $A_{\lambda} \sim \lambda^p$.  They found a good fit to the heavily extinguished SN 2006X if LMC-like grains, described by a power $p=-2.5$, are confined within a circumstellar shell of $R_{CSM} = 10^{16} \, \mathrm{cm}$ \, \citep{Goobar_Scat}.  By comparison, their simulations show that Galactic dust follows a shallower extinction law with $p=-1.5$.

Due to its proximity, SN 2014J was well studied in UV-optical-near IR (NIR) wavelengths and the wavelength dependence of the extinction law has been analyzed by a number of groups.  \citet{Goobar} and \citet{Amanullah_2014J}, using similar optical and near-infrared photometry, determined that a simple (one parameter) reddening law characterized by $R_V\sim1.4$ due to $A_V \sim 1.7-1.9\, \mathrm{mag}$ of extinction could adequately match their data.  This was achieved by artificially reddening SN 2011fe, which was not thought to be affected by much extinction, with a Milky Way-like extinction law while allowing E(B-V) and $R_V$ to vary.  Equally consistent with the \citet{Amanullah_2014J} dataset was a power-law model with an index $p=-2.1$.  The photometric and spectroscopic measurements of \citet{Foley} were also inconsistent with a reddening law characterized by $R_V = 3.1$.  Additionally, they found that a multiple-component extinction model gave an overall better fit to their spectroscopic data of SN 2014J than with either a simple reddening or circumstellar model alone.  Their best fit model included interstellar dust with properties ($R_V=2.59 \pm 0.02$) similar to some lines of sight in the Milky Way \citep{CCM} and a circumstellar component that scattered light similar to the LMC.  However, \citet{Johansson} did not find an IR excess in 3.6 $\mathrm{\mu m}$ and 4.5 $\mathrm{\mu m}$ \emph{Spitzer} data which placed limits on $M_{dust} < 10^{-5} \, \mathrm{M_{\odot}}$ within $10^{17}$ cm of the explosion and thus argued against the CSM scattering models.  Optical and UV photometry obtained with \emph{Swift} also conclude that SN 2014J's extinction is due to interstellar dust since the light curves are well matched by a reddened SN 2011fe template without the need for invoking CSM scattering \citep{Brown}.  Similar results were found with an extended dataset between $0.2-2 \,\mathrm{\mu m}$ \citep{Amanullah_diversity}.

An independent measure of host galaxy extinction is possible through high resolution absorption spectra which reveal weak interstellar features.  Atomic and molecular lines as well as diffuse interstellar bands (DIBs) are found in the spectrum of SN 2014J at velocities that arise within the gas of M82 \citep{Welty}.  The equivalent width of the DIB $\lambda 5780$  \AA, in particular, has been shown to be a proxy for host galaxy ISM reddening \citep[and references within]{Phillips_Na}.  The strength of this DIB along the line of sight to SN 2014J gives an estimate of $A_V \sim 1.95 \, \mathrm{mag}$ for the visual extinction \citep{Welty}.  Therefore, nearly all of the extinction measured by SN 2014J photometry can be accounted for with interstellar dust.

Temporal variations detected in some high resolution spectra of SNe Ia have been proposed as evidence of circumstellar medium (CSM) near the supernova explosion \citep{Patat_CSM}.  The width and intensity of the Na I doublet lines of SN 2006X were found to vary with time between -2 and +61 days after maximum light.  The authors proposed that the evolution in the line profiles was caused by the ionization of the CSM by UV radiation of the supernova which slowly recombines with time.  The gas near the supernova which undergoes ionization could be due to successive nova eruptions or from the stellar wind of a red giant companion star.  Variability of the Na I lines does not seem to be a common property of SNe Ia because only 3 of 17 SNe Ia with multi-epoch high-resolution spectra have shown changes in the equivalent width (EW) of their Na I lines \citep{Simon, Patat_CSM, Dilday, Sternberg_2014}. Variability was also detected in SN 1999cl using low-resolution spectra \citep{Blondin_CSM}.  The ability to detect these variations, however, depends on the window of observations and the geometry of the CSM.  \citet{Foley}, who had the most complete sample of high-resolution spectra in the weeks around maximum light, did not see any changes in the Na I lines of SN 2014J between -10 and +18 days after the peak brightness.  \citet{Maeda_ISM} confirmed these results with a dataset that extends to 255 days after maximum and find that all of the Na D absorptions arise in interstellar dust $\sim 40$ pc in the foreground.   

The detection of a light echo near the explosion sites of some SNe Ia is also considered to be evidence of CSM.  Whether CSM light echoes have been detected is the subject of current debate.  SN 1998bu was shown to have two echoes at distances $<10$ pc and at 120 pc from the supernova; the former has been argued to be created by circumstellar dust \citep{Garnavich}.  A more extended echo was present in SN 2006X leading to estimates of a cloud of dust located $27 - 170$ pc \citep{Wang_LE06X}.  This was hinted to at least have a partial local origin, although see \citet{Crotts_LE06X} for an alternate analysis.  Light echoes in SN 2014J were first discovered by \citet{CrottsLE} with further analysis of the light echo expansion presented by \citet{Yang_Echo}.  The authors find a luminous inner arc created by a foreground dust sheet at 222 pc and a faint outer ring consistent with foreground dust at 367 pc.  Multiple inner echoes are also seen at smaller radii indicating a complex ISM within M82.

Optical and NIR broadband polarization of SN 2014J showed a degree of polarization which decreased quickly from shorter to longer wavelengths with a peak in polarization at a much shorter wavelength than typically measured by Galactic stars \citep{Kawabata}.  The authors proposed that the majority of the polarization was due to interstellar dust in M82 since the polarization angle was nearly constant across the spectrum.  The wavelength dependence of the broadband measurements was better fit with an inverse power law than a Serkowski curve which underestimated the IR polarization by 0.2-0.3\%.  They concluded that this unusual behavior suggests the polarizing dust grains of M82 have mean radii $< 0.1 \,\mathrm{\mu m}$.  These results were echoed by \citet{Patat_Dust} using spectropolarimetry of SN 2014J at day +1.  

Here we present multiple epochs of optical spectropolarimetry of SN 2014J from maximum light to 111 days past-maximum with the paper organized as follows.  In Section 2 we present the details of our observations and data reduction process.  In Section 3, we discuss our estimate of the ISP and its implications for the dust of M82.  We discuss the ISP subtracted intrinsic polarization of SN 2014J in Section 4.  In Section 5 we explore how the polarization of SN 2014J compares to other supernovae before summarizing our conclusions in Section 6.

\section{Observations}

We observed SN 2014J over seven nights at the Steward Observatory 2.3-m Bok (Kitt Peak, AZ) and 6.5-m MMT (Mt. Hopkins, AZ) telescopes with the CCD Imaging/Spectropolarimeter \citep[SPOL]{SPOL}.  A rotatable semi-achromatic half-waveplate positioned just below the slit acts as a retarder.  A Wollaston prism located in the collimated beam separates the orthogonally polarized (ordinary and extraordinary) components onto a 800$\times$1200 pixel CCD. 

Obtaining the linear Stokes parameters for each target requires two $Q$ and two $U$ images with each image recording the ordinary and extraordinary spectra.  For each image, the total exposure time is split between four waveplate rotation angles that are separated by 90 degrees.  Each of the four rotations produces an identical polarization state on the detector and the result reduces the effects of variations in the waveplate as a function of rotator angle.  The second $Q$ or $U$ image is obtained with waveplate positions offset by 45 degrees thereby swapping the ordinary and extraordinary beams on the detector so that pixel-to-pixel variations can be minimized during the reduction.  The sequence of exposures which produces the four Stokes images is repeated several times for each target to increase the signal-to-noise.

We report the observation log in Table \ref{table1}, where the supernova phase is determined with respect to \emph{B}-band maximum light which occurred on 2014 February 2 \citep{Foley}.  Our 7 nights of observations were grouped into 6 epochs.  The polarization did not change significantly between May 22 and 24 so these nights were combined into a single epoch.

We reduced the data using custom, but mature IRAF\footnote{IRAF is distributed by the National Optical Astronomy Observatory, which is operated by the Association of Universities for Research in Astronomy (AURA) under a cooperative agreement with the National Science Foundation.} routines.  We began by bias-subtracting and flat-fielding each image in the usual manner and used observations of He, Ne, and Ar lamps at the beginning of each run for wavelength calibration purposes.  We then extracted the ordinary and extraordinary traces and used the 1-D spectra to measure the linear Stokes parameters $Q$ and $U$.  We de-biased the positive definite nature of the polarization calculation using the prescription $P = \pm \sqrt{\mid Q^2 + U^2 - \frac{1}{2}(\sigma^2_Q - \sigma^2_U)\mid}$ \citep{Wardle} where the sign is determined according to the sign of the modulus.  Finally, to further increase the signal-to-noise ratio, we binned the final Stokes parameters, after removal of the ISP, to 20  \AA ~or 60  \AA ~wide intervals.

During each run, we obtained multiple observations of the polarized standards BD+59 389 (2014 February 2 and 9), HD245310 (2014 February 2, 9, 24 and March 25), HD154445 (2014 February 25), HD161056 (2014 February 25), VI Cyg \#12 (2014 February 25 and April 20), and Hiltner 960 (2014 February 25 and April 20) to determine the linear polarization position angle on the sky \citep{Schmidt}.  We used the average position angle offset from these stars to correct the spectra from the instrumental to the standard equatorial frame.  Additionally, we confirmed that the instrumental polarization is less than 0.1\% by observing the unpolarized standard stars G191-B2B (2014 February 2, 9, 25, March 25, April 20) and BD+28$^{\circ}$4211 (2014 April 20).

\section{Interstellar Polarization}
The wavelength dependence of continuum polarization detected along the line of sight to SN 2014J may be due to the combination of the interstellar medium (ISM) in the Milky Way, ISM in the host galaxy, or dust local to the supernova.  In the sections following, we discuss how pure transmission through dust or a combination of transmission and scattering by dust manifests itself in the spectropolarimetric data and we discuss the implications of each scenario.  For more details, please see Appendix \ref{appendix}.

\begin{figure}
\includegraphics[angle=90,width=\linewidth]{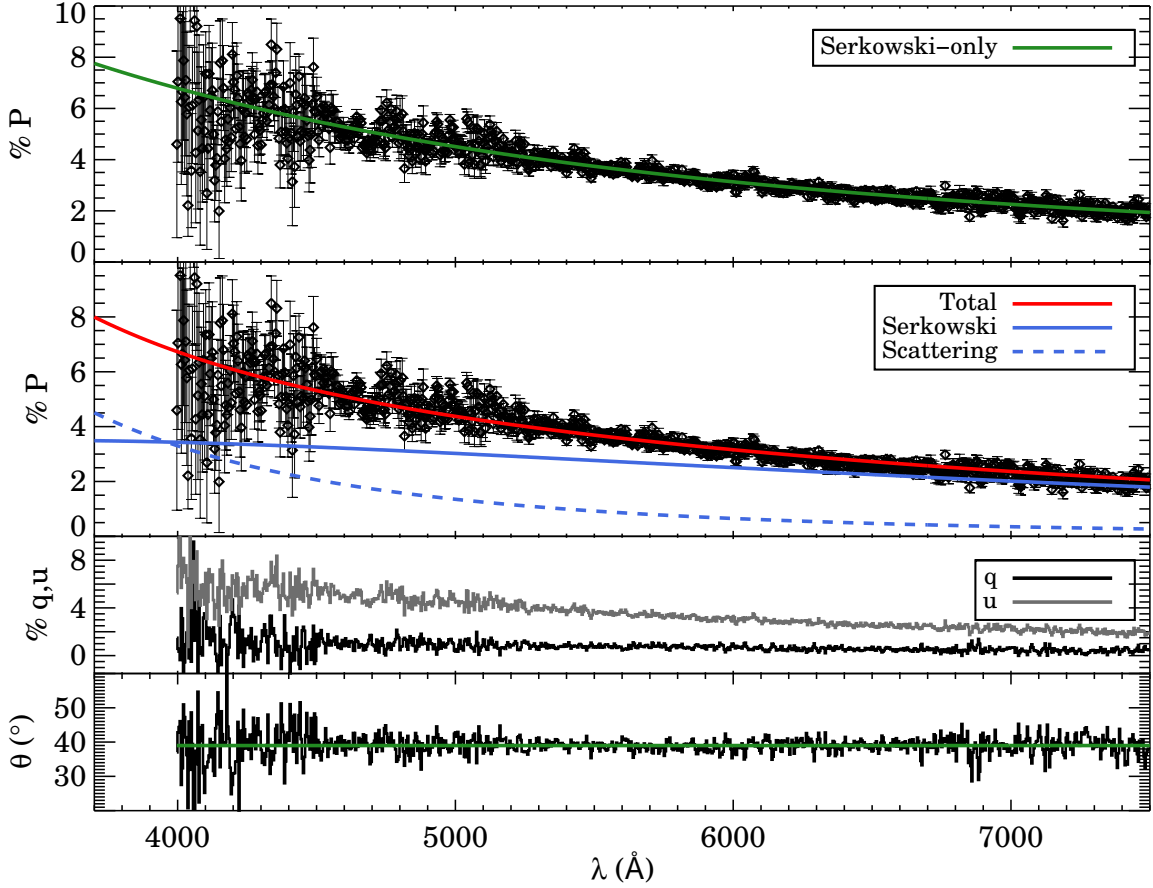}
\caption{
The observed degree of polarization at +51 days, binned to 4  \AA, with our best fit Serkowski-only curve (solid green), assuming that all of the continuum polarization is due to the ISP of M82, is displayed in the top panel (see Sec. \ref{Serkowski}).  The upper-middle panel shows how a Serkowski curve (solid blue) in addition to a scattering component (dashed blue) compare to the Epoch 4 data (see Sec. \ref{Combo}).  The sum of the two components is shown in solid red.  The $q$ (black) and $u$ (gray) spectra are shown in the lower-middle panel and the polarization angle is presented in the lower panel.  The average value of 39$^{\circ}$, found between $5000- 7500$  \AA, is shown in green.
\label{isp}
}
\end{figure}

\begin{figure}
\includegraphics[angle=90, width=\linewidth]{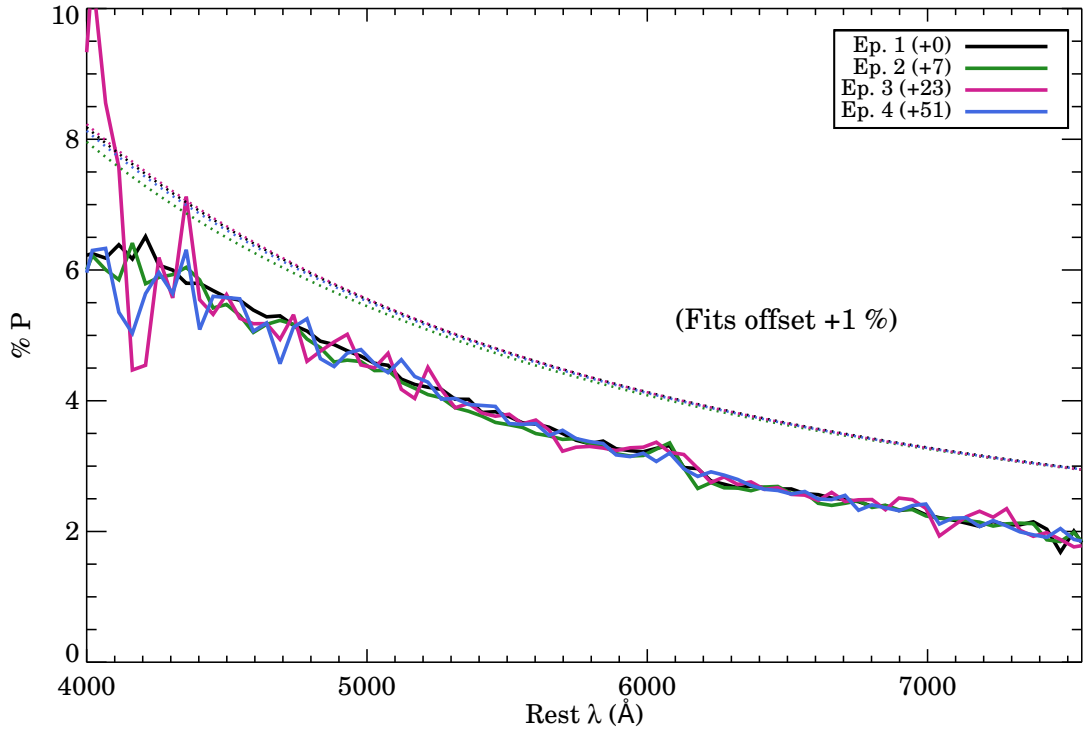}
\caption{Continuum polarization prior to ISP subtraction is binned to 47 \AA ~ and shown for Epoch 1 (black), Epoch 2 (green), Epoch 3 (pink), Epoch 4 (blue).  To check for variability, we applied a least-squares power-law fit of the form $P(\lambda) = c \, (\lambda/0.55 \, \mathrm{\mu m})^{-\beta}$ to each epoch.  These fits are shown as dotted lines and have been offset by 1\%  for clarity.
}
\label{continuum}
\end{figure}

\subsection{Transmission through dust}
\label{Serkowski}
To determine the intrinsic polarization of the supernova's ejected layers, we must first derive the contribution from aligned dust grains within the ISM, known as ISP.  We present the polarization of Epoch 4, 51 days after maximum light, in Fig. \ref{isp}.  The average continuum polarization declines monotonically from $\sim7\%$ at 4000  \AA ~to $\sim2\%$ at 7500 \AA.  Earlier spectra which are closer to maximum light show a very similar level of continuum polarization, but have a strong modulation near 6150 \AA ~, as seen in Fig. \ref{continuum}, which is associated with line polarization intrinsic to the supernova.  Later spectra at +77 , +104, and +106 days, have a lower signal-to-noise ratio than Epoch 4, so we do not use them to determine the level of ISP.  

Both the Milky Way and the supernova's host galaxy may contribute an ISP component to the observed continuum polarization.  There are no Galactic stars within a 2$^{\circ}$ box centered on the supernova with polarization measurements  to the best of our knowledge \citep{Heiles}.  However, the highest degree of polarization measured for a star at a similar Galactic latitude as M82 is 0.76\% with 0.24\% being the average among 17 stars with comparable latitudes. Therefore, we conclude the Milky Way's contribution to the ISP is small in comparison to the polarization along the line of sight to SN 2014J.  \citet{Hoffman} note that due to the vector nature of polarization, even if the Milky Way ISP is small compared with that of the supernova, its contribution may cause a position angle rotation if the vector alignment is unfavorable.  Using the ratio of the highest degree of polarization for a Galactic star noted above to the average polarization observed near 4500 \AA ~in Epoch 4, we find that the intrinsic polarization angle of SN 2014J could differ by as much as $3.9^{\circ}$ from what we observe \citep[Eqn. 1]{Hoffman}.

Our approach to estimating the host galaxy's ISP contribution is to wait until the ejecta has become optically thin to electron scattering during the supernova's nebular phase.  At this point any residual polarization is expected to be contributed by intervening dust.  Because the spectrum shows a sharply decreasing degree of polarization with wavelength, a single value of $q$ and $u$ cannot properly describe the ISP.  Instead the wavelength dependence of the polarization spectrum, if due completely to the linear dichroism of interstellar dust, can be captured by fitting the data with the empirical Serkowski relation 

\begin{equation} \label{Serk}
P_{ISP}(\lambda) = P_{max} \, \exp \left(-K\, \ln^2 \, \frac{\lambda_{max}}{\lambda}\right)
\end{equation}    

\noindent where $\lambda_{max}$ is the wavelength at which the polarization reaches a maximum, $P_{max}$, while $K$ controls the width of the curve \citep{Serkowski1975}.  The relationship was determined from the optical polarization of Galactic stars and typical values are $\lambda_{max}=0.55 \,\mathrm{\mu m}$ and a curve width dependent on $\lambda_{max}$ such that $K=0.01 \pm 0.05 + (1.66 \pm 0.9) \, \lambda_{max}$ where $\lambda_{max}$ is given in $\mathrm{\mu m}$ \citep{Whittet1992}.  Because the peak in the degree of polarization of SN 2014J appears to be $< 0.4 \, \mathrm{\mu m}$ as evident in Fig.\ref{isp}, the derived parameters will likely be peculiar.  Our least-squares fitting routine yields  $K=0.40 \pm 0.03$ \footnote{Note, we do not force $K$ to conform to the Whittet relation.}, $\lambda_{max}=0.0456 \pm 0.1052 \,\mathrm{\mu m }$, and $P_{max}=45.04 \pm 30.56\%$.  Our values agree well with the results of \citet{Patat_Dust} even though we use a polarization spectrum obtained 50 days later.  However, our results are not as easily compared to the broadband measurements, as \citet{Kawabata} held $K$ fixed to fit their polarization, which was  averaged over 5 epochs between -11 and +33 days.

It has been proposed that the wavelength dependence of the polarization is related to the extinction law through the relation $R_V \simeq 5.5\lambda_{max}$ \citep{Serkowski1975}.  The low $\lambda_{max}$ value derived from polarization implies a low $R_V$ consistent with photometric and spectroscopic estimates \citep{Foley}.  However, near-UV (NUV) and optical photometry of M82 show evidence of a 2175 \AA ~bump in the galaxy's NUV color-color plots, suggesting that this particular galaxy may be more comparable to the Milky Way than other starburst galaxies which often do not show the bump in their extinction laws \citep{Hutton_2014}.  In a further analysis of this photometry, which was taken in the years before the explosion of SN 2014J, the dust at a projected distance of 1 kpc from the nucleus, the same distance as the supernova \citep{Foley}, has an $R_V \sim 3-4$ \citep{Hutton_2015}.  This is obviously in contrast to the low values of $R_V$  measured after the explosion \citep{Foley,Goobar,Amanullah_2014J}.  However, we note that Fig. 5 in \citet{Hutton_2015} displays a large range in selective extinction values near this distance.

\subsection{Combined scattering and transmission through dust}
\label{Combo}

Light that encounters dust located at interstellar distances from the supernova acquires polarization through linear dichroism.  The polarization angle in this case aligns with the magnetic field direction of the galaxy in which the grains are aligned.  The polarization signal of scattering dust in the circumstellar environment, however, will be due to photons scattered into the line of sight.  The perpendicular electric field vectors for dust distributed symmetrically about the line of sight would cancel completely, similar to an unresolved symmetric supernova envelope \citep{Hoflich}.  Therefore, to measure a net linear polarization, there must be some asymmetry in the dust's overall geometry which will be reflected in the associated polarization angle.

To demonstrate how circumstellar dust may be present in their polarization spectrum, \citet{Patat_Dust} applied a multi-component fit to their day +1 spectropolarimetric data of SN 2014J composed of a Serkowski curve for interstellar dust and a Rayleigh scattering component which dominates at short wavelengths.  They find a respectable fit using $K=1.15$ (held fixed), $\lambda_{max}=0.35\,\mathrm{\mu m}$ and $P(\lambda)=P_{s}(0.4 \,\mathrm{\mu m} / \lambda)^4$ with $P_{s}=0.8P_{max}$.  From their Fig. 6, we estimate $P_{max}\sim3.8\%$ which gives $P_s=3.0\%$ for the scattering component.  Additionally, scattering contributes $\sim 40 \%$  of the total polarization at $0.4 \,\mathrm{\mu m}$. 

Because \citet{Patat_Dust} use a spectrum near maximum light, their fit may include a small level of intrinsic continuum polarization attributed to electron scattering in the supernova ejecta.  Later observations are less likely to have an intrinsic supernova component, so we present a similar fit to our +51 day epoch in Fig. \ref{isp}.  We find that our data are well described with $\chi^2=1.45$ using $K=1.15$, $\lambda_{max} = 0.35 \, \mathrm{\mu m}$, $P_{max} = 3.5\%$, and $P_s = 3.3\%$ with scattering dominating at wavelengths less than $0.395 \, \mathrm{\mu m}$.  The scattering also contributes about 56\% of the total polarization in our fit.  We chose to hold K and $\lambda_{max}$ fixed to decrease the number of free parameters by using the original K value presented by \citet{Serkowski1975} and $\lambda_{max}$ chosen to represent one of the bluest peaks observed in the Milky Way (See Appendix \ref{appendix}).  It is important to note that our fit in Fig. \ref{isp} is just for illustrative purposes as a number of decompositions could potentially describe the observed polarization.  For comparison, the multi-component fit of \citet{Patat_Dust} gives a $\chi^2=2.42$ to our Epoch 4 data and is even worse, $\chi^2=7.01$, when compared to our maximum light epoch.

Additionally, we find an average angle of $\theta=39^{\circ} \pm 1.78$ between 5000-7000 \AA ~in Epoch 4, which is in agreement with the earlier-epoch polarization measurements of \citet{Kawabata} and \citet{Patat_Dust}.  \citet{Greaves} determined that a $40^{\circ}$ angle describes the magnetic field structure of M82, thus suggesting that if there is circumstellar dust polarizing light by scattering, the dust must be oriented nearly identical to the dust lanes of M82 \citep{Patat_Dust}.  \citet{Hoang} discussed a toy model that may be able to explain the lack of change in angle.  If small grains which undergo Rayleigh scattering are trapped within an accretion disk aligned in the magnetic equatorial direction of the white dwarf and the magnetosphere of the star contains a size distribution of grains reminiscent of the Galaxy, the polarization vectors that emerge from the two grain populations will be parallel.  However, \citet{Hoang} mentions that these dust grains will likely be swept away within $1-2$ hours after the explosion, depending on the ejecta velocity, and therefore are not expected to be observable at the epochs of polarization we discuss here.  We note that because the contribution of the small Milky Way ISP may rotate the observed polarization angle by as much as $3.9^{\circ}$ from its intrinsic value (see Sec. \ref{Serkowski}), it makes it more likely that such an apparent alignment between the circumstellar dust of the supernova and the ISP of M82 could occur by simple chance.

\subsection{Stokes ISP Estimate}
We have presented both the results of fitting a Serkowski function that assumes transmission through dust of a single distribution and the results of a two-component model that includes transmission through dust and a scattering component.  However, because there are concerns with adopting either scenario, we also present a low order polynomial fit to the rotated Stokes parameters (RSP) of $q_{RSP}$ and $u_{RSP}$ of Epoch 4.  We use the fits to determine the non-varying ISP and in turn the supernova's intrinsic polarization.  We note that this prescription for determining the ISP from a late epoch makes use of the assumption that the continuum polarization is null as a result of the ejecta that is optically thin to electron scattering.  

The rotated parameters are derived by revolving the $q- u$ plane through an angle $\theta$ such that all of the continuum polarization then lies along $q_{RSP}$ while $u_{RSP}$ represents deviations from the dominant axis \citep{TrammellRSP,LeonardSP1999em}.  Presenting the supernova's intrinsic $P$ as the rotated quantities allows us to avoid the high level of bias that accompanies the traditional determination of $P$ through $P=\sqrt{q^2+u^2}$ and also avoid the problems caused by the asymmetric error distribution of $P$ for low signal-to-noise data in the debiased formula  $P = \pm \sqrt{\mid Q^2 + U^2 - \frac{1}{2}(\sigma^2_Q + \sigma^2_U)\mid}$ \citep{Wardle}.

To calculate the RSP, we use the form $q_{RSP} = q \, \cos \, 2\theta + u \, \sin\,2\theta$ and $u_{RSP} = -q \, \sin\,2\theta + u \, \cos \, 2\theta$ and use the mean polarization angle derived between $5000- 7500$ \AA  ~in Epoch 4 as our $\theta$ since the angle is nearly constant over all epochs.  We then subtract the following best-fit polynomials from the rotated observed Stokes vectors

\begin{eqnarray} \label{ISPStokes}
	q_{ISP} &=& 4.57 - 17.10\,(\lambda-0.5\mathrm{\mu m}) + 27.10\,(\lambda-0.5\mathrm{\mu m})^2 \nonumber\\
	u_{ISP} &=& -0.004 - 0.008\,(\lambda-0.5\mathrm{\mu m})
\end{eqnarray}

\noindent and the remainder is considered to be intrinsic to the supernova.  The statistical uncertainties on the coefficients of the $q_{ISP}$ polynomial are respectively, 0.02, 0.28, and 1.12 while the uncertainties on the  $u_{ISP}$ fit are 0.013 and 0.090.

We binned the observed polarization as high as 47 \AA ~to check for any variability in the continuum polarization among the various epochs and find none as shown in Fig. \ref{continuum}.  We quantify the lack of variability in two ways.  First, we fit a power-law of the form $P(\lambda) = c \, (\lambda/0.55 \,\mathrm{\mu m})^{-\beta}$ to the spectrum of each epoch prior to ISP subtraction and find the fits are in good agreement with each other.  Additionally, the best-fit $K-\lambda_{max}$ we derive from fitting the Serkowski curve to each epoch lie within the $4-\sigma$ confidence contour drawn relative to Epoch 4's best-fit (see Fig. \ref{confidence}).  Although there is some scatter in the best-fit $K-\lambda_{max}$ with time, the degree of polarization defined by these curves at 5500 \AA ~all agree within $\sim 0.01\%$.  If circumstellar dust is present at close enough distances to the supernova, we expect the polarization to change as the dust is evaporated by radiation which would in turn change the grain size distribution as small grains disappear.  Calculations by \citet{Amanullah_CSM} find that sublimation will deplete dust that resides at $r \le 10^{16} \mathrm{cm}$ which is nearly equivalent to 4 light-days.  This is prior to all of our spectropolarimetric observations so it is not surprising that we do not see variability in the degree of polarization among our epochs.  Therefore, we assume the ISP remains constant and does not change with time allowing the same ISP correction to be applied to each epoch of observation.

\section{Intrinsic polarization}

\begin{figure}
\includegraphics[angle=90,width=\linewidth]{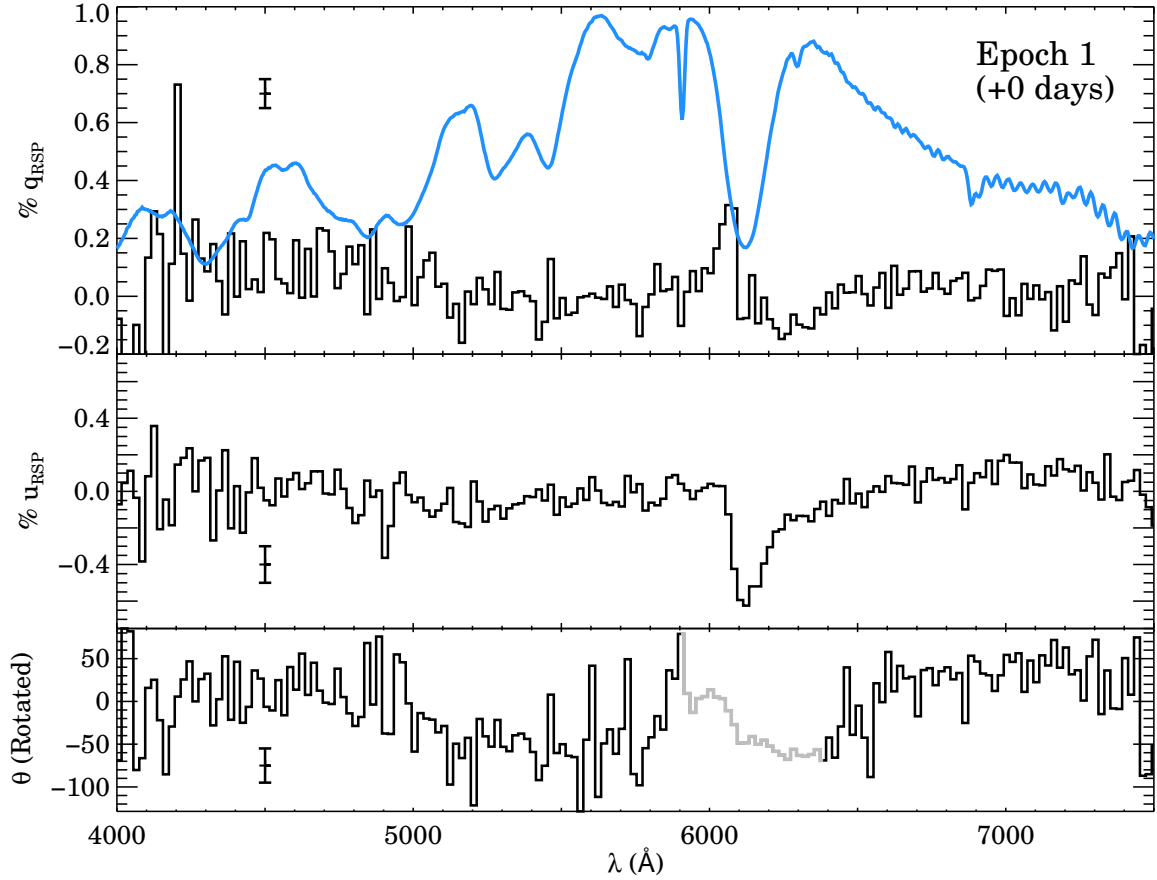}
\caption{
The rotated Stokes parameters after correction for ISP are shown for 2014 February 2 (Epoch 1) with the arbitrarily scaled flux spectrum (bold blue line) in the upper panel.  Below the rotated polarization spectra is the rotated angle.  The polarization angle near the Si II 6355 \AA ~feature is highlighted over the wavelengths $5900- 6400$ \AA.  All spectra are in the rest-frame of the host galaxy and binned to 20 \AA.  The average 1$- \sigma$ error bars for the polarization and position angle spectra are displayed at 4500 \AA. 
}
 \label{Ep1Pol}
\end{figure}


\begin{figure}
\includegraphics[angle=90, width=\linewidth]{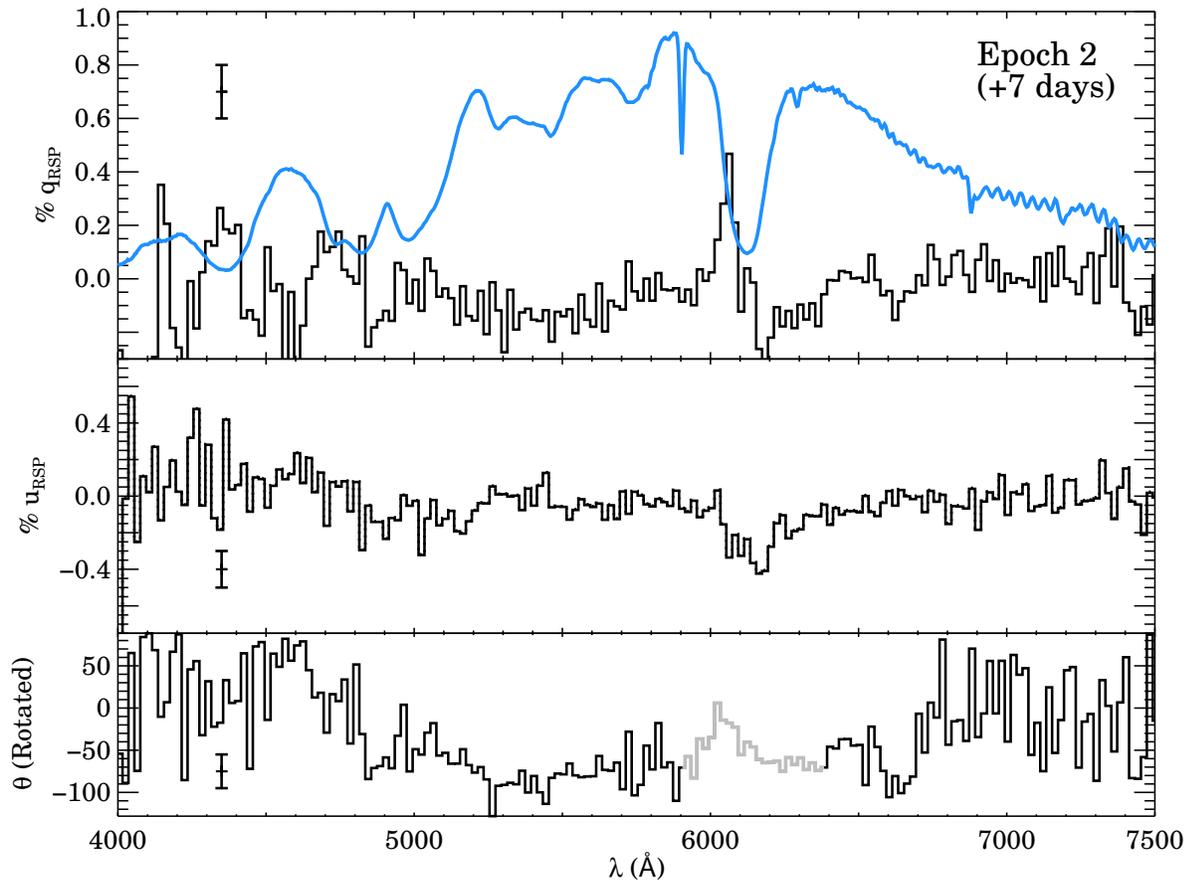}
\caption{Same as Fig. \ref{Ep1Pol} for 2014 February 9 (Epoch 2), but with the 1-$\sigma$ error bars displayed at 4350 \AA.  
\label{Ep2Pol}
}
\end{figure}


\begin{figure}
\includegraphics[angle=90, width=\linewidth]{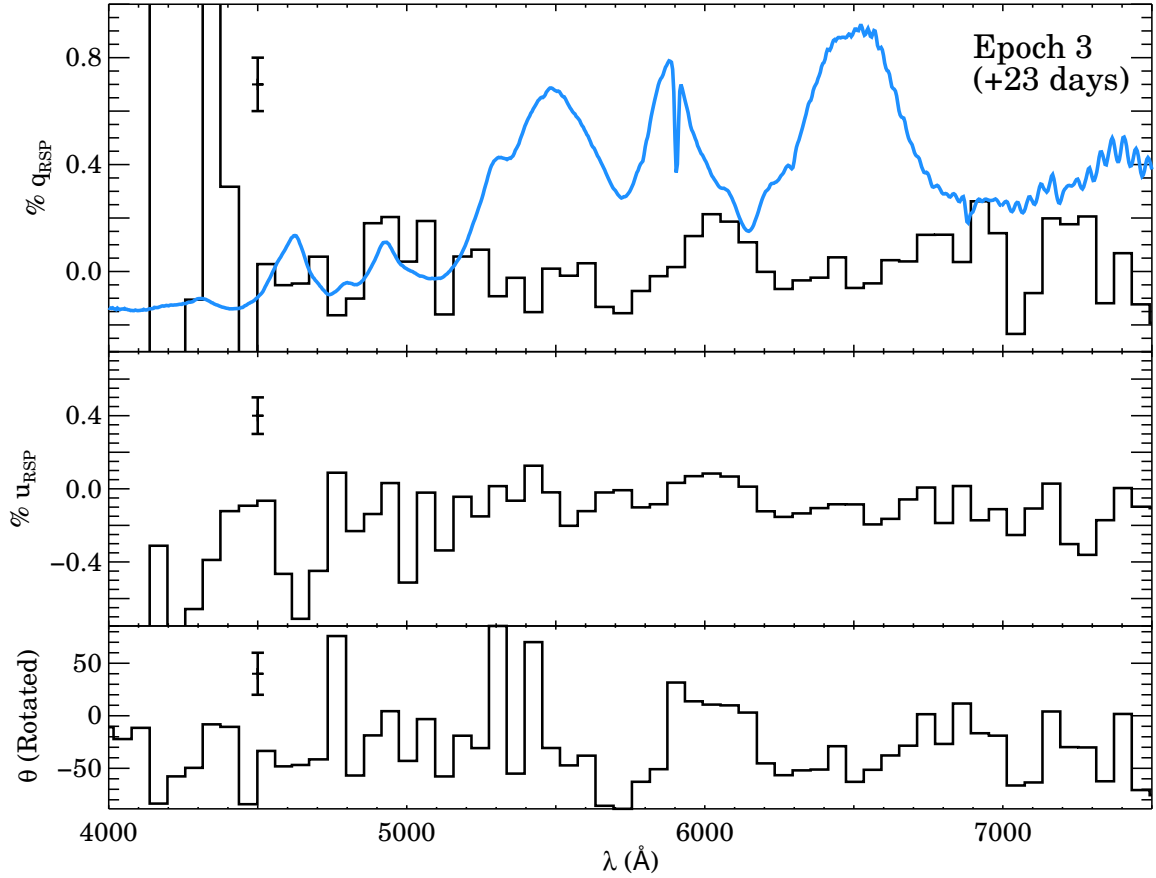}
\caption{Same as Fig. \ref{Ep1Pol} for 2014 February 25 (Epoch 3), except all spectra are binned to 60 \AA.  Note the $q_{RSP}$ axis has been plotted over a different range from earlier epochs.  The polarization angle is not highlighted across the absorption line at 6100 \AA ~as there is no clear line polarization visible at the same wavelengths.  The average 1-$\sigma$ error bars are displayed at 4500 \AA.
\label{Ep3Pol}
}
\end{figure}

\begin{figure}
\includegraphics[angle=90, width=\linewidth]{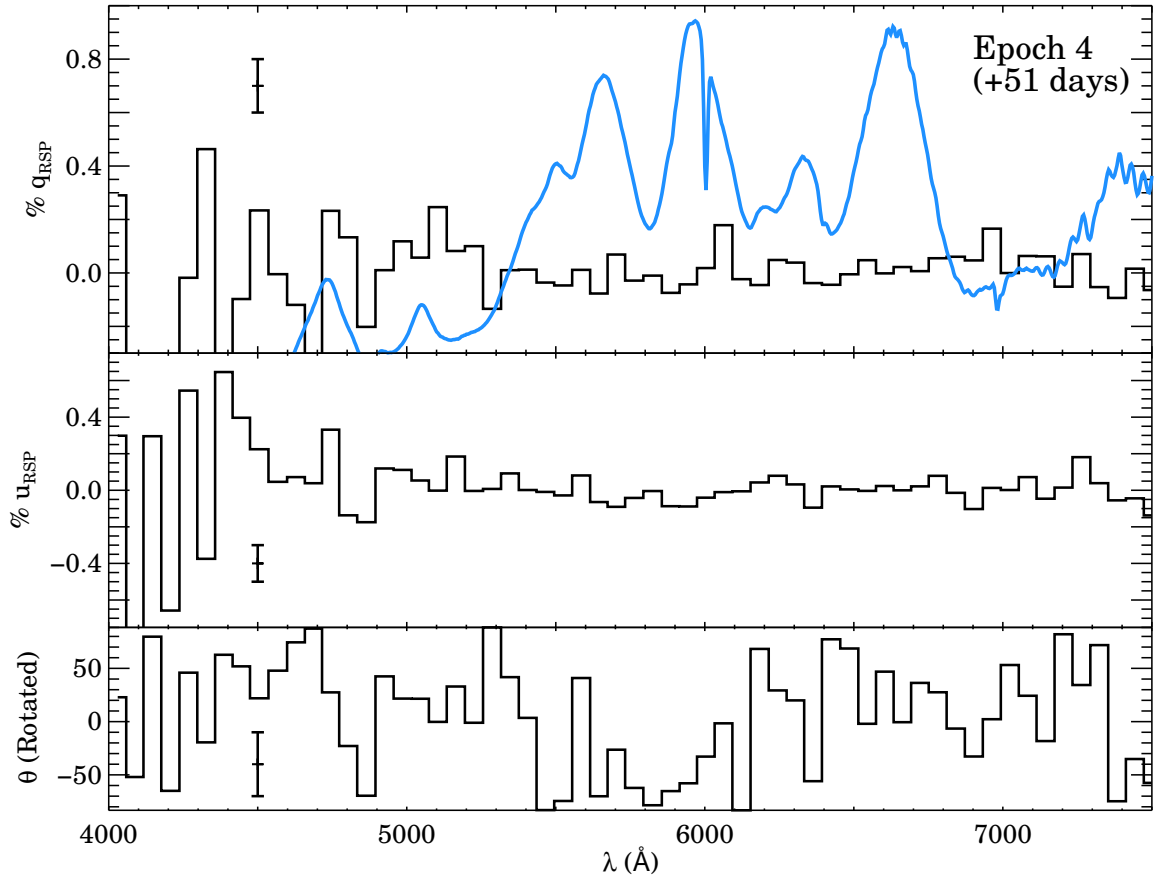}
\caption{Same as Fig. \ref{Ep3Pol} for 2014 March 25 (Epoch 4), with all spectra binned to 60 \AA.  The average 1-$\sigma$ error bars are displayed at 4500 \AA.
\label{Ep4Pol}
}
\end{figure}

\begin{figure}
\includegraphics[angle=90, width=\linewidth]{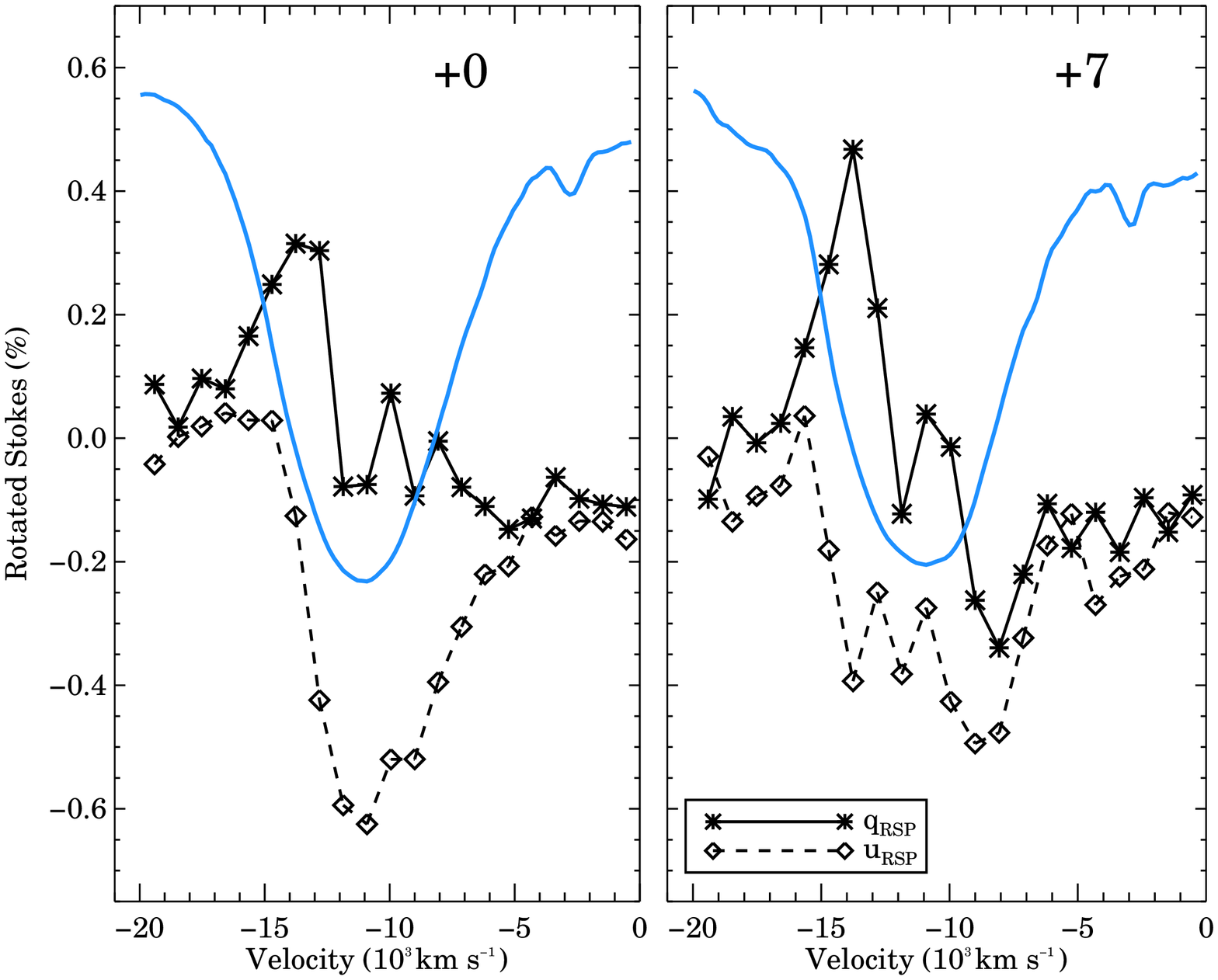}
\caption{Variation in the level of polarization with velocity over the Si II 6355 \AA ~profile between 5900 and 6400 \AA ~for days +0 and +7.  The smooth blue curve traces the arbitrarily scaled flux spectrum, the $q_{RSP}$ spectrum is displayed as stars connected by a solid line, and the $u_{RSP}$ spectrum is shown as open diamonds connected by a dashed line.  All polarization spectra are in the rest frame of the host galaxy and binned to 20 \AA.
\label{VelPol}
}
\end{figure}

After removal of the interstellar dust component which was chosen by fitting polynomials to our late-time Epoch 4 spectra, the supernova's intrinsic degree of polarization remains.  We present SN 2014J's intrinsic spectropolarimetric evolution over the first four epochs in Figs. \ref{Ep1Pol}- \ref{Ep4Pol}.  Our results show that the Si II line 6355 \AA ~of the supernova is indeed intrinsically polarized near maximum light.  We discuss this in detail below.  It is common for SNe Ia to have a conspicuous feature in this line \citep{SPReview} and SN 2014J is no exception.  Often the NIR Ca II triplet near 8000 \AA ~is polarized as well, but our wavelength coverage does not reach this far into the red so we cannot comment on this characteristic. 

\subsection{Continuum polarization} \label{secContinuum}

Assuming that the continuum polarization is defined by a region of the spectrum that is free of strong lines, we measure the average continuum between 6600-7400 \AA ~ in Epochs 1-4 to have an intrinsic polarization of $0.09 \pm 0.07\%$, $0.01 \pm 0.17\%$, $0.14 \pm 0.17\%$, and $0.03 \pm 0.07\%$, respectively.  Therefore, all epochs are consistent with zero polarization.  This low value is in agreement with other normal SNe Ia measurements which typically show $< 0.3\%$ polarized continua light \citep{SPReview}.  The subluminous SNe Ia SN 1999by and SN 2005ke and the normal SN 2011fe have shown higher detections of 0.5-0.7\% \citep{SP1999by,SP2005ke,SP2011fe}.  The small degree of continuum polarization indicates that globally the photosphere is nearly spherically symmetric and is consistent with less than 10\% deviation from a perfect sphere \citep{Hoflich}.

\subsection{Si II polarization}
\label{Intrinsic}

Near maximum light, changes in the polarization degree are clearly detected at the Si II 6355 \AA ~line.  Fig. \ref{VelPol} shows the first two epochs' degree of polarization in velocity space.  In Epoch 1, the polarization in $q_{RSP}$ rises to a peak near $-14,000 \,\mathrm{km \, s^{-1}}$.  Furthermore, a broad depression in the $u_{RSP}$ spectrum forms from  $-14,500 \,\mathrm{km \, s^{-1}}$ to $-4000 \,\mathrm{km \,s^{-1}}$ across the Si II line.  Meanwhile, the continuum has an average $u_{RSP}$ value of $0.08 \pm 0.07\%$ at this epoch.  We use the $u_{RSP}$ spectrum to quote a polarization measurement for Si II at this epoch and find an average $P_{SiII} = 0.54 \pm 0.05\%$ between $-13,500 \,\mathrm{km \, s^{-1}}$ and $-9,500 \,\mathrm{km \,s^{-1}}$.  We also note the Si II polarization peaks near the line's absorption minimum which we measure to be at $-11,000 \,\mathrm{km \, s^{-1}}$.

In the following epoch at +7 days, the peak in $q_{RSP}$ has increased to $0.46\%$ at  $-14,000 \,\mathrm{km \, s^{-1}}$, but this value decreases to $0.32 \pm 0.13\%$ if averaged over  $-15,000 \,\mathrm{km \, s^{-1}}$ to $-12,000 \,\mathrm{km \,s^{-1}}$.  Meanwhile, the feature in $u_{RSP}$ has grown more complex with three dips present over the same wavelength range as Epoch 1.  We measure a peak polarization level of $0.37 \pm 0.05\%$ between $-10,000 \,\mathrm{km \, s^{-1}}$ and $-7000 \,\mathrm{km \, s^{-1}}$.

The importance of obtaining spectropolarimetric data to learn details about the ejecta distribution is showcased by Fig. \ref{VelPol}.  The flux profile of the Si II line shows only minor changes between the first and second epochs, however, the polarization profiles have evolved significantly.  \citet{McCall} predicted that the polarization profile should peak at an absorption line's minimum since the absorbing ejecta blocks the direct, unpolarized light from the photosphere along the line of sight.  We see this in Epoch 1 where the peak in $u_{RSP}$ occurs nearly in alignment with the absorption trough.  With the evolution in the polarization profile in Epoch 2, we see that the asymmetry in the line has changed with the line depth in the ejecta as the fastest and the slowest material are now nearly just as polarized as the line center.  The absorption trough's polarization has also decreased significantly between the two epochs from $0.63\%$ to $0.26\%$ at the same velocity slice indicating that the more direct light from the photosphere is not being absorbed as readily as a week prior.

In the last two epochs analyzed at +23 and +51 days, the Si II feature is no longer detected in the polarization spectra (see Fig. \ref{Ep3Pol} and \ref{Ep4Pol}).  By +23 days, the Si II 6355 \AA ~feature is still present in the flux spectrum, but Fe II has begun encroaching on the wings and a combination of Fe II, Co II, and Cu II are present between 6100-6800 \AA ~at +51 days \citep{Vallely}.  With the lack of Si II in both the flux and polarization spectra, it is not surprising that the polarization angle no longer shows smooth changes across this region of the spectrum.  In general, the angle has become more erratic because the error in polarization angle is very large when the degree of polarization is nearly zero.

Overall, the spectropolarimetric evolution of SN 2014J is similar to previous SNe Ia observations and theory \citep{SPReview}.  Line polarization is typically visible near maximum light and often at levels less than 1\%.  Earlier measurements showed line polarization that peaked before maximum, however SN 2014J joins a growing sample of SNe Ia that peak later (see Fig. \ref{Comparison}).  In the weeks following maximum light, the photosphere recedes deeper into more uniform ejecta or the regions of asymmetric material become optically thin causing the line polarization to fade with time.  This often occurs within two weeks after the light curve peak, but one SN Ia has shown re-polarization of the Ca II NIR lines nearly 40 days after maximum \citep{SP2006X}.  \citet{SP2007sr} also report levels of 2-4\% for the Ca II NIR triplet of SN 2007sr at +63 days.  Earlier measurements are not available, so it is not clear whether this is a re-polarization of the line.

\begin{figure}
	\includegraphics[angle=90, width=\linewidth]{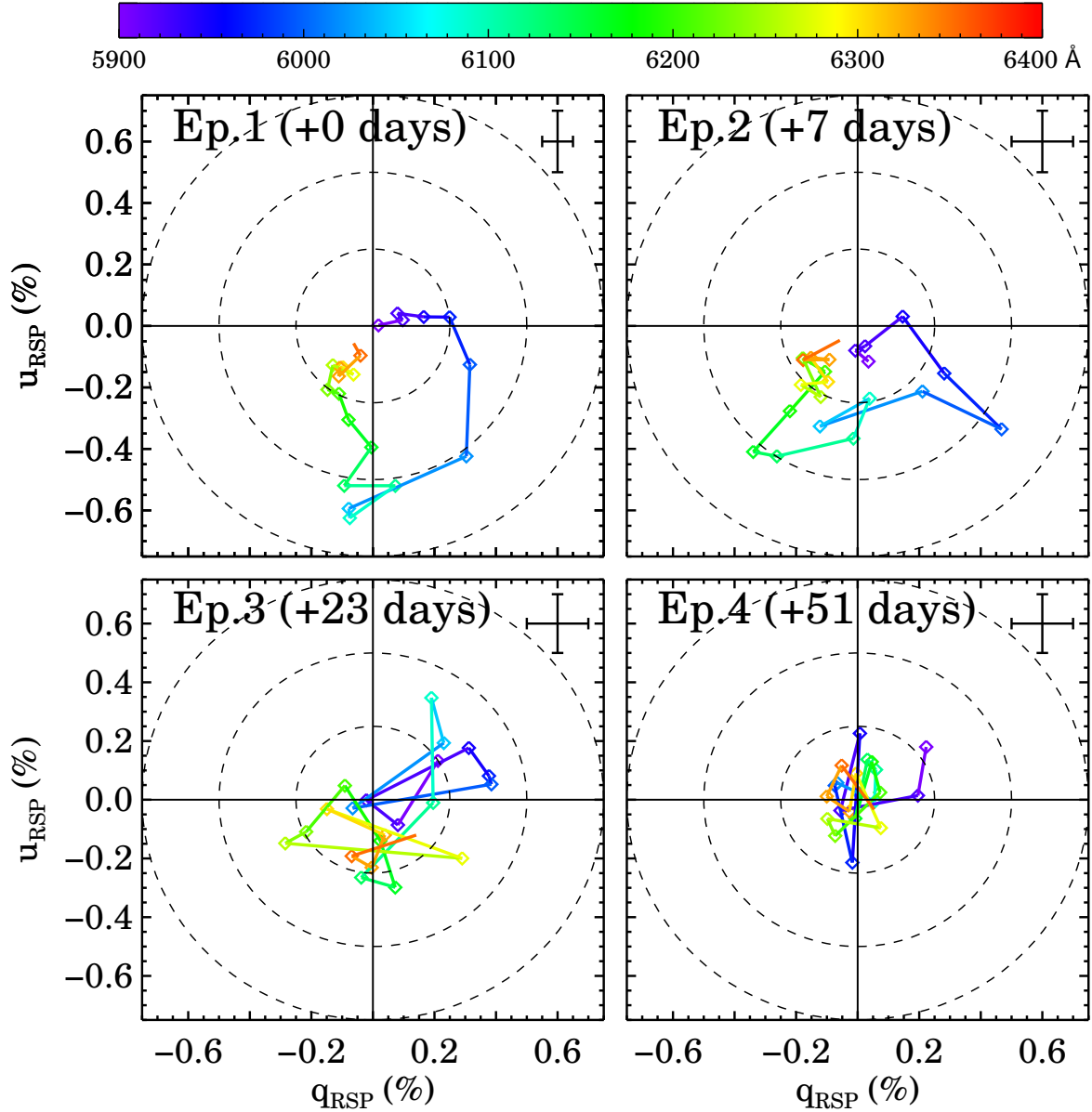}
	\caption{ISP-corrected $q_{RSP}- u_{RSP}$ diagrams of Si II 6355 \AA , binned to 20 \AA, over the first four epochs of observation color-coded according to wavelength.  Polarization levels of 0.25, 0.5 and 0.75\% are indicated by dashed lines and the average 1-$\sigma$ error-bar size is shown in the top right of each panel.}
	\label{SiQU}
\end{figure}

Although the polarized spectra are useful for comparing the levels of polarization of different SNe Ia, investigating the data on the $q- u$ plane is often more enlightening.  Fig. \ref{SiQU} shows the $q_{RSP}$ and $u_{RSP}$ polarization for the range 5900-6400 \AA , which includes the Si II absorption line.  Each data point corresponds to a $q_{RSP}$-$u_{RSP}$ vector pair with the wavelength represented by the symbol's color.  In the first epoch, the Si II feature forms a loop as the $u_{RSP}$ polarization steadily increases from the blue towards the middle of the line and then begins to decrease towards the red.  The middle of the absorption trough shows the highest degree of polarization, extending just beyond the 0.5\% level depicted as a dashed line.  Moving to the second epoch, the complex nature of the $u_{RSP}$ spectrum can be seen here as well as three loops form past the $0.25\%$ polarization circle.  Finally, by the third and fourth epochs, when we are unable to definitely identify the Si II polarization, the loop behavior has diminished as well.

Loops in the $q$-$u$ plane occur as the polarization angle rotates on the sky.  If the distribution of Si II were axisymmetric, the $q_{RSP}- u_{RSP}$ vector pairs would arrange themselves in a straight line on the plane.  The presence of a loop, rather than a linear collection of points, suggests that the distribution of Si II is not axisymmetric throughout the ejecta.  Similar loop features have been found in a number of SNe Ia such as SN 2001el \citep{SP2001el}, SN 2004S \citep{SP2004S}, SN 2009dc \citep{SP2009dc} and SN 2011fe \citep{SP2011fe}.  \citet{Kasen} used parametrized models to explore possible origins of the HV Ca II loop of SN 2001el.  Either an ellipsoidal shell rotated with respect to the photosphere or large-scale clumpy matter could create the change in polarization angle.  Both of these scenarios would partially block polarized light of the underlying continuum leading to line polarization.

\section{Discussion}

\begin{figure}
	\includegraphics[angle=90, width=\linewidth]{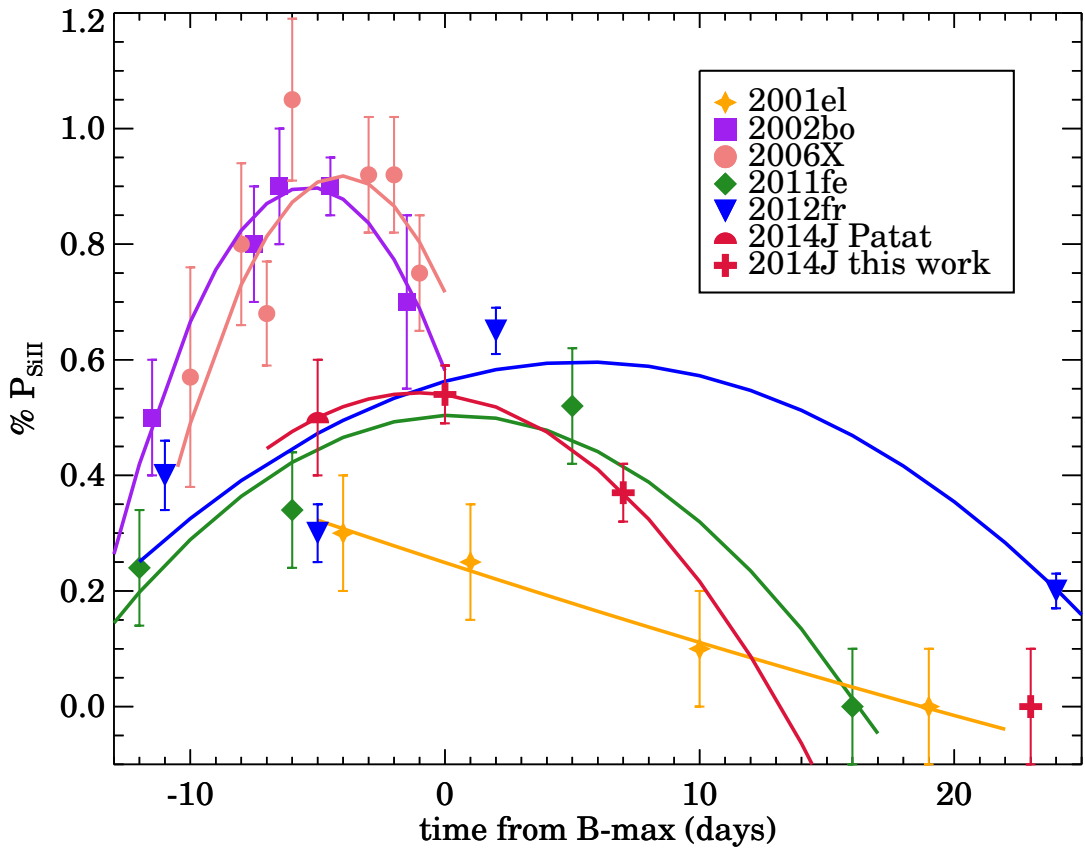}
	\caption{Comparison of the Si II 6355 \AA ~line polarization among SNe Ia with multiple epochs of observations.  A weighted least squares polynomial or  linear fit  has been added to demonstrate the temporal evolution of each supernova's Si II polarization.  The polarization of SN 2001el (yellow stars) was estimated from \citet{SP2001el}.  SN 2002bo (purple squares) is from \citet{Wang_Sci}, SN 2006X (pink circles) data are from \citet{SP2006X} and SN 2011fe (green diamonds) is from \citet{SP2011fe}.  The SN 2012fr (blue upside-down triangles) observations of \citet{SP2012fr} showed both high velocity and photospheric components.  The high velocity component dominated at -11 days, but was diminished by the second epoch when the photospheric polarization became more prominent.  Preliminary data (red half-circle) of \citet{Patat_ATEL}, in addition to the measurements of this work (red crosses) are coupled together to show the evolution of SN 2014J's Si II polarization.  Our null measurement at +23 days, has not been included in the fit for SN 2014J.}
	\label{Comparison}
\end{figure}

It is clear from our spectropolarimetric results that SN 2014J is intrinsically polarized.  The overall asphericity of the electron-scattering photosphere is small, but significant undulations across the Si II line around maximum light indicate that there are deviations from this geometry.  The changes in line polarization over time in SN 2014J also showcase the importance of obtaining multiple epochs of observation.  This allows a detailed study of the element distribution in the outer layers of SNe Ia ejecta.

Although the sample of SNe Ia with polarization measurements has been growing since the first evidence of an intrinsically polarized SN Ia was obtained with SN 1996X \citep{SP1996X}, only a handful of these studies are multi-epoch.  In an effort to compare SNe Ia at the same epoch, \citet{Wang_Sci} fit the Si II polarization of 17 explosions, each with a single pre-maximum measurement, with a second-order polynomial to determine the time dependence of the line polarization.  This polynomial was then used to correct a supernova's measured polarization to -5 days.  Using such a model seemed appropriate at the time because the evolution of SN 2002bo, determined over 7 epochs, matched this derived polynomial well.  However, SN 2014J does not seem to fit this trend.  Using the +0 day measurement for SN 2014J, a Si II polarization of 0.66\% is expected at -5 days according to the temporal evolution polynomial.  This value is quite a bit higher than the value of 0.5\% measured on the same day by \citet{Patat_ATEL}.  

In Fig. \ref{Comparison}, we present the time dependence of Si II 6355 \AA ~polarization for all well-studied SNe Ia.  The measurements of SN 2002bo, SN 2006X, SN 2011fe, and SN 2012fr are adapted directly from figures, tables or the text of their publications \citep{Wang_Sci,SP2006X,SP2011fe, SP2012fr}.  Meanwhile, we estimate the values for SN 2001el from its polarization spectra \citep{SP2001el}.  We also included the preliminary value for SN 2014J on day -5 as presented by \citet{Patat_ATEL}.  We fit each supernova's time dependence with a least-squares polynomial or linear line (for SN 2001el) which is also displayed in Fig. \ref{Comparison}.  For the SN 2014J fit, we used the -5 day value from \citet{Patat_ATEL}, +0, and +7 day values from this work, but did not include our +23 day measurement since it is consistent with 0\%.  We use the peak value of $u_{RSP}$ (see Sec.\ref{Intrinsic}) for SN 2014J as this seems most consistent with the literature.  In addition, the fit of SN 2012fr's temporal evolution involves a mixture of high-velocity (at -11 days) and photospheric velocity polarization components.                  

We notice several trends when comparing the temporal evolution of these supernovae.  First, there are differences in the level and time of peak polarization among SNe Ia.  Both SN 2002bo and SN 2006X display very similar evolution in their Si II polarization, which reached a level of  $\sim$1\% polarization approximately 5 days prior to maximum light.  More recent supernovae, however, such as SN 2011fe and SN 2012fr, reach a lower polarized peak of 0.5-0.6\%, but only at maximum light or later.  SN 2014J appears to have a similar evolution.  Unfortunately, SN 2001el only shows declining levels of polarization, so it is hard to determine how it compares to the other supernovae.

Second, there is variety in the length of time a supernova shows asymmetries in its Si distribution.  We see that the polarization of SN 2002bo and SN 2006X rises and falls rapidly whereas SN 2011fe, SN 2012fr, and SN 2014J are more gradual.             

Finally, in addition to displaying the highest levels of Si II polarization, SN 2002bo and SN 2006X also exhibited high expansion velocities.  Their photospheric velocity as measured by the absorption of Si II was in excess of $-13,000 \,\mathrm{km\,s^{-1}}$ at maximum light \citep{Silverman}.  Thus, these two explosions are listed as HV in the \citet{Wang_golden} classification and HVG (``high velocity gradient") by \citet{Benetti}.  Meanwhile, SN 2001el and SN 2011fe belong to the NV (``normal velocity") and LVG groups as they had maximum light velocities $\sim-10,000\,\mathrm{km\,s^{-1}}$ \citep{Silverman}.  SN 2012fr and SN 2014J, however, are not easily sub-classified by photospheric velocity.  \citet{Childress} measured the maximum light velocity of SN 2012fr to be $\sim -12,000 \,\mathrm{km\,s^{-1}}$ which places this supernova just over the dividing line into the HV category.  A long plateau in the Si II expansion, however, kept the photospheric velocity at -12,000 km $\mathrm{s}^{-1}$ for nearly six weeks, classifying SN 2012fr as LVG.  Likewise, SN 2014J is categorized as HV and LVG (see Sec.\ref{PreviousObs}).     

In combining these three trends, we see that the HVG supernovae SN 2002bo and SN 2006X reach the highest levels of polarization recorded in SNe Ia and on much earlier timescales than their LVG counterparts.  Therefore, these explosions show larger degrees of asphericity that are located in higher layers of ejecta as compared to LVG supernovae since the photosphere is deeper within the ejecta when LVG explosions show their highest degree of polarization.  As a direct comparison, SN 2006X's photospheric velocity, as measured by the Si II 6355 \AA ~line, was over -17,000 $\mathrm{km\,s^{-1}}$ near its peak in polarization \citep{SP2006X} while SN 2014J's velocity was -11,000 $\mathrm{km\,s^{-1}}$.

Although a relationship between the temporal evolution of a supernova's asymmetries and its ejecta velocity does seem possible, the sample size of well-observed SNe Ia is still small, so we caution against overinterpretation.  However, this is not the first time a correlation between these two observables has been suggested \citep{Leonard}.  \citet{Maund_uni} discovered a linear correlation between Si II polarization and the supernova's velocity gradient characterized by the best-fit line $P_{SiII} = 0.267 + 0.006\,\dot{v}_{SiII}$.  The HVG explosions exhibited more asymmetry than the LVG when comparing the level of Si II polarization measured at, or corrected to, -5 days using the formula determined by \citet{Wang_Sci}.  We find that SN 2014J also follows this relationship.  We expect a polarization level of 0.52\% using the velocity gradient determined by \citet{Marion} which is similar to the preliminary value of 0.5\% presented by \citet{Patat_ATEL}.    

It has been speculated that the luminosity of SNe Ia may be related to asymmetry as well.  \citet{Wang_Sci} observed a relationship between the polarization of a supernova and its brightness as measured by the decline rate parameter.  The observed decline rate for SN 2014J as measured by \citet{Foley}, predicts a Si II polarization of 0.28\%, which is too low for the -5 day value measured by \citet{Patat_ATEL}.  However, it has been shown that reddening decreases the observed decline rate \citep{Phillips_Color}.  Correcting the supernova's brightness for reddening increases the expected polarization to $0.36- 0.45\%$ using $\Delta m_{15}(B)_{\mathrm{true}}=1.01- 1.08$ \citep{Foley}, which at the upper end is closer to reproducing SN 2014J's -5 day Si II polarization. 

In conclusion, we find that SN 2014J has a similar evolution as other LVG supernovae, which show moderate levels of peak polarization at maximum light or later, in contrast to HVG explosions.  It is therefore not surprising that SN 2014J's -5 days polarization measurement cannot be reproduced with the temporal evolution polynomial of \citet{Wang_Sci} which only matches the Si II time dependence of HVG supernovae.  We therefore advise extra caution when using this equation for supernovae without multiple epochs of observation.  It may be more worthwhile to estimate a supernova's pre-maximum polarization if multiple epochs are not available using the velocity gradient relationship which estimated a similar degree of polarization as was actually measured for SN 2014J at -5 days.  Although, the sample size is small, our Fig. \ref{Comparison} reflects the main conclusion of the velocity gradient correlation which is that HVG supernovae have higher levels of line polarization.

\section{Summary}
The spectropolarimetry presented here reveals that the heavily reddened, but otherwise normal Type Ia SN 2014J is polarized near maximum light, allowing us to explore the multi-dimensional nature of the explosion.  A consistently low detection of continuum polarization implies a nearly symmetric explosion, at least from our viewing angle, but higher levels across the Si II 6355 \AA ~line point to a more complicated structure in the ejecta above the photosphere.

The continuum polarization of SN 2014J, estimated between 6600-7400 \AA,  is statistically consistent with zero between +0 and +51 days.  However, the line polarization reaches levels of 0.54\% on day +0 and 0.37\% on day +7, measured as the average degree of polarization over a small velocity range.  At maximum light, the minimum of the absorption trough shows the highest polarization at 0.63\%.  A week later, however, the same velocity slice has decreased in polarization by $\sim 0.4\%$.  The polarization feature at this epoch has also evolved into several peaks spread out over the center of the absorption line that are nearly equivalent in degree of polarization.  The changes in the polarization profile between the first two epochs indicates that the line's asymmetry is evolving with the line depth within the ejecta.  Despite these large fluctuations in the degree of polarization, we see minimal changes in the flux profile.  By +23 and +51 days, Si II 6355 \AA ~only makes a small contribution to the flux spectrum and we no longer detect line polarization, indicating that the asymmetry diminishes with time or depth in the ejecta.  Another interesting aspect of the Si II polarization is the loop it forms on the $q_{RSP}- u_{RSP}$ diagram.  This type of structure develops when the polarization angle varies monotonically across a line.  A skewed ellipsoidal shell of the ion or perhaps clumps of Si II caused by anisotropies in the $^{56}$Ni excitation are possible explanations.  The evolution of SN 2014J's polarization is similar to other LVG supernovae which tend to reach a maximum level of asymmetry around maximum light or later.  Supernovae with higher expansion velocities reach their peak polarization approximately 5 days earlier.  LVG explosions overall also show lower degrees of polarization at their peaks than their HVG counterparts.     

Although the intrinsic polarization of SN 2014J appears normal relative to other SNe Ia, the ISP is more remarkable.  It has been suggested that at least part of the extinction along the line of sight is due to scattering by circumstellar dust.  We find the polarization angle aligns closely with the dust lanes of M82 and the degree of polarization does not change over our period of observation, suggesting that the continuum polarization has an interstellar origin.  However, the derived parameters from our Serkowski fit are unusual and may not be physical.  The wavelength dependence of the continuum polarization also provides an independent measure of $R_V$ that does not require supernova colors.  The blue wavelength at which the Serkowski curve estimates the peak degree of polarization invokes an abundance of very small aligned dust grains in the ISM of M82 compared to the Milky Way and a steep extinction law characterized by $R_V < 2$.  However, UV and optical photometry of the galaxy derive a higher value.  Our polarization data are also well described by the combination of a Serkowski curve and a scattering component where scattering contributes nearly half of the polarization at 4000 \AA.  This interpretation results in ISM dust grains that are still smaller than in the Milky Way, but less extreme than in the pure ISP model.  However, we see a lack of variability in the wavelength dependence of the polarization angle.  Variability is expected when the dust of the two components have different orientations so, if this model is credible, the two grain populations must share a similar orientation.  Therefore, we cannot distinguish between a pure ISP or a combination of ISP and CSM dust at this time, but it is likely that there is an enhanced abundance of small grains along the line of sight to SN 2014J.

\begin{deluxetable}{ccccccccccc}
\setlength{\tabcolsep}{0.035in} 
\tabletypesize{\scriptsize}
\tablecaption{Observation log of SN 2014J \label{table1}}
\tablewidth{0pt}
\tablehead{
\colhead{UT Date} & \colhead{$\mathrm{Days^a}$} & \colhead{Epoch} & \colhead{$\mathrm{Exposure^b}$} & \colhead{Telescope} & \colhead{Wavelength} & \colhead{Grism} & \colhead{Slit} & \colhead{Dispersion} & \colhead{Spatial scale} & \colhead{Resolution}
\\
& (+$B_{max}$) & & ($sec$) &  &( \AA) & $(lines\,mm^{-1}$) & $(arcsec.)$ & ( \AA$\,pixel^{-1}$) & $(pixel\,arcsec.^{-1}$) & ( \AA)
}
\startdata
	2014 Feb 2.37    & +0     & 1 &   3$\times$160 &   Bok    & 4000-7500	& 600   & 5.1  &  4.14  & 1.9  & 32    	\\
	2014 Feb 9.38    & +7     & 2 &   3$\times$160 &   Bok    & 4000-7500  & 600   & 4.1  &  4.14   & 1.9  & 26	\\
	2014 Feb 25.40  & +23    & 3 &   4$\times$240 &   Bok   & 4000-7500	& 600   & 4.1  &  4.14  &  1.9  & 26	\\
	2014 Mar 25.40  & +51    & 4 &   3$\times$360  &  Bok   & 4000-7500  & 600   & 4.1  &  4.14  & 1.9  & 26 	\\
	2014 Apr 20.36   & +77   & 5 &   1$\times$480 &   MMT & 4050-7200 	& 964   & 1.9  &  2.62  & 5.2  & 20	\\
	2014 May 22.16  & +109 & 6 &   2$\times$480 &   Bok   & 4000-7500   & 600   & 4.1  & 4.14  &  1.9  & 26	\\	
	2014 May 24.16  & +111 & 6 &   2$\times$480 &   Bok    & 4000-7500	 & 600  & 4.1  & 4.14  & 1.9  &  26 	 \\
\enddata
\tablenotetext{a}{Days are calculated with respect to maximum brightness in the B band which occurred on 2014 February 2 \citep{Foley}}
\tablenotetext{b}{Exposure times listed are those required to obtain one Stokes parameter}
\end{deluxetable}

\acknowledgements
We thank the anonymous referee for their careful reading and constructive comments which improved the presentation of our work.  This research has made use of the NASA/IPAC Extragalactic Database (NED), which is operated by the Jet Propulsion Laboratory, California Institute of Technology, under contract with the National Aeronautics and Space Administration and the VizieR catalogue access tool, CDS, Strasbourg, France.  DCL acknowledges support from NSF grants AST-1009571 and AST-1210311, under which part of this research was carried out.  This research was also supported by NSF grants AST-1210599 (UA) and AST- 1210372 (U. Denver).  Some observations reported here were obtained at the MMT Observatory, a joint facility of the University of Arizona and the Smithsonian Institution."

\appendix
\section{Wavelength dependence of polarization }
\label{appendix}

\setcounter{figure}{0} \renewcommand{\thefigure}{A.\arabic{figure}}

\begin{figure}
\includegraphics[angle=90, width=\linewidth]{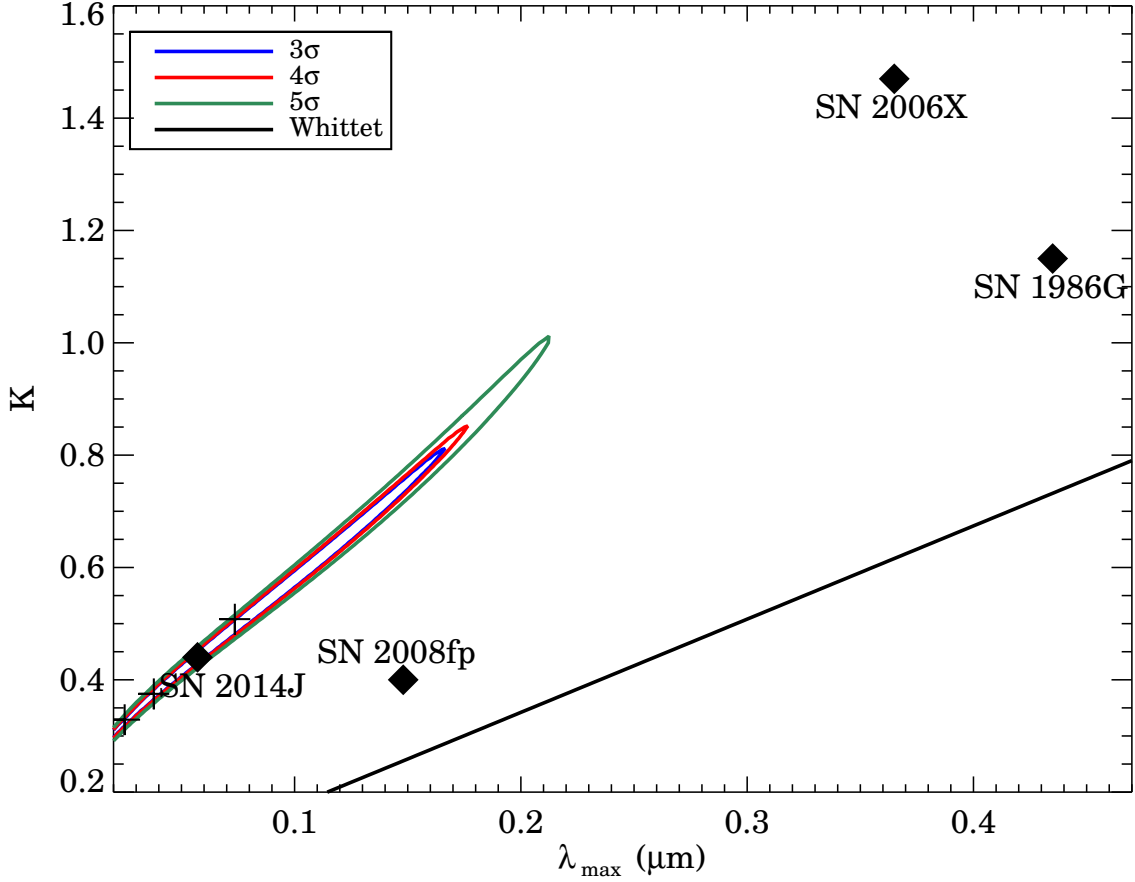}
\caption{Contours depicting the 3-, 4-, and 5-$\sigma$ confidence levels indicate how K depends on $\lambda_{max}$ for our Serkowski-only fit to SN 2014J.  Best fit values for SNe 2014J (Epoch 4, this work) 2008fp \citep{SP2008fp}, 2006X \citep{SP2006X}, and 1986G \citep{SP1986G} are shown as black diamonds while the Whittet relationship $K=0.01+1.66\lambda_{max}$ that describes Galactic stars \citep{Whittet1992} is shown as a solid black line.  The cross-marks indicate the best-fit $K-\lambda_{max}$ values from (bottom-left to top-right) Epoch 3, Epoch 2, and Epoch 1.
}
\label{confidence}
\end{figure}

Although we can impose a Serkowski curve on the continuum polarization's wavelength dependence, the ability to accurately pinpoint the parameters is a cause for concern.  We created a $\chi^2$ map of the $\lambda_{max}$-K plane around our best fit Serkowski values given in Sec. \ref{Serkowski}.  The confidence level contours in Fig. \ref{confidence}, drawn relative to the minimum $\chi^2$, shows the 3-5$\sigma$ levels.  The diagram supports the idea that an increase in K should follow an increase in the peak polarization to redder wavelengths as given by the Whittet relationship, but more generally the extended contours express that a number of $\lambda_{max}$-K pairs is applicable.  Restated another way, the parameters cannot be uniquely determined when the peak in polarization is outside the observed range.  Comparing the contours to the Whittet relationship shown as a solid black line in the plot indicates that SN 2014J's Serkowski parameters lie quite far from those expected in the Galaxy.  Also shown as crosses on the plot are the best-fit $K-\lambda_{max}$ pairs from earlier epochs.  Because the level of continuum polarization does not show variability between +0 and +51 days, the earlier epochs lie along the extended contours and less than $4-\sigma$ from the Epoch 4 parameters.

Also, the derived Serkowski parameters of SN 2014J differ from those determined for Galactic stars.  Out of the 104 lines of sight sampled by \citet{Whittet1992}, only 5 have a $\lambda_{max} < 0.4 \,\mathrm{\mu m}$, 4 of which are within the Cygnus OB2 clustering of young stars obscured by the Cygnus Rift.  One such example with a blue $\lambda_{max}$ is Cygnus OB2 No.12 which exhibited $P_{max}=9.90\pm0.67\%$, $K=0.60\pm0.08$, and $\lambda_{max}=0.35\pm0.005 \, \mathrm{\mu m}$.  Thus, it is rare for Galactic stars to show peaks in polarization outside of optical wavelengths.  By contrast, several highly reddened SNe Ia\textemdash1986G \citep{SP1986G}, 2006X \citep{SP2006X}, and 2008fp \citep{SP2008fp}\textemdash display atypical wavelength dependencies and seem to deviate from the Whittet relation by at least 3 or more standard deviations \citep[see][Fig. 3]{Patat_Dust}.  Similar to SN 2014J, both SN 2006X and SN 2008fp showed $\lambda_{max} < 0.4 \, \mathrm{\mu m}$.  Additionally, it is not clear whether the Serkowski curve can be safely extrapolated to ultraviolet wavelengths.  For the small number of stars with adequate UV polarization measurements \citep{UV_1999}, the Serkowski relation provides a decent fit, unlike in the infrared which is better expressed by a power law.  However, in some cases for $\lambda < 0.28 \, \mathrm{\mu m}$, the Serkowski relation underestimates the observed polarization \citep{UV_1992, AndersonUVPol}.     

Assuming that the Serkowski method is valid for ISP along the line of sight to SN 2014J, we deduce that the dust properties of M82 differ from the Galaxy since the best-fit values are outside the expected ranges from stellar probes.  The Mie theory for dielectric grains estimates the size, $a$, of the average polarizing dust grain as $\lambda_{max}=2\pi \big(n-1\big) a$ \citep{Whittet_book}.  Under the presumption that silicates are the main contributors to the polarization as in the Milky Way \citep{Voshchinnikov}, then a refractive index $n=1.6$ would imply a grain radius of $a=0.015 \, \mathrm{\mu m}$ in M82 in contrast to the Galactic $a=0.15 \, \mathrm{\mu m}$.  Hence, an overabundance of small, aligned grains is implied within M82.

\bibliographystyle{apj}
\bibliography{specpol_bib}

\begin{thebibliography}{}
\expandafter\ifx\csname natexlab\endcsname\relax\def\natexlab#1{#1}\fi

\bibitem[{{Amanullah} \& {Goobar}(2011)}]{Amanullah_CSM}
{Amanullah}, R., \& {Goobar}, A. 2011, \apj, 735, 20

\bibitem[{{Amanullah} {et~al.}(2014){Amanullah}, {Goobar}, {Johansson},
  {Banerjee}, {Venkataraman}, {Joshi}, {Ashok}, {Cao}, {Kasliwal}, {Kulkarni},
  {Nugent}, {Petrushevska}, \& {Stanishev}}]{Amanullah_2014J}
{Amanullah}, R., {Goobar}, A., {Johansson}, J., {et~al.} 2014, \apjl, 788, L21

\bibitem[{{Amanullah} {et~al.}(2015){Amanullah}, {Johansson}, {Goobar},
  {Ferretti}, {Papadogiannakis}, {Petrushevska}, {Brown}, {Cao}, {Contreras},
  {Dahle}, {Elias-Rosa}, {Fynbo}, {Gorosabel}, {Guaita}, {Hangard}, {Howell},
  {Hsiao}, {Kankare}, {Kasliwal}, {Leloudas}, {Lundqvist}, {Mattila}, {Nugent},
  {Phillips}, {Sandberg}, {Stanishev}, {Sullivan}, {Taddia}, {{\"O}stlin},
  {Asadi}, {Herrero-Illana}, {Jensen}, {Karhunen}, {Lazarevic}, {Varenius},
  {Santos}, {Sridhar}, {Wallstr{\"o}m}, \& {Wiegert}}]{Amanullah_diversity}
{Amanullah}, R., {Johansson}, J., {Goobar}, A., {et~al.} 2015, \mnras, 453,
  3300

\bibitem[{{Anderson} {et~al.}(1996){Anderson}, {Weitenbeck}, {Code},
  {Nordsieck}, {Meade}, {Babler}, {Zellner}, {Bjorkman}, {Fox}, {Johnson},
  {Sanders}, {Lupie}, \& {Edgar}}]{AndersonUVPol}
{Anderson}, C.~M., {Weitenbeck}, A.~J., {Code}, A.~D., {et~al.} 1996, \aj, 112,
  2726

\bibitem[{{Benetti} {et~al.}(2005){Benetti}, {Cappellaro}, {Mazzali},
  {Turatto}, {Altavilla}, {Bufano}, {Elias-Rosa}, {Kotak}, {Pignata}, {Salvo},
  \& {Stanishev}}]{Benetti}
{Benetti}, S., {Cappellaro}, E., {Mazzali}, P.~A., {et~al.} 2005, \apj, 623,
  1011

\bibitem[{{Blondin} {et~al.}(2009){Blondin}, {Prieto}, {Patat}, {Challis},
  {Hicken}, {Kirshner}, {Matheson}, \& {Modjaz}}]{Blondin_CSM}
{Blondin}, S., {Prieto}, J.~L., {Patat}, F., {et~al.} 2009, \apj, 693, 207

\bibitem[{{Brown} {et~al.}(2015){Brown}, {Smitka}, {Wang}, {Breeveld}, {de
  Pasquale}, {Hartmann}, {Krisciunas}, {Kuin}, {Milne}, {Page}, \&
  {Siegel}}]{Brown}
{Brown}, P.~J., {Smitka}, M.~T., {Wang}, L., {et~al.} 2015, \apj, 805, 74

\bibitem[{{Bulla} {et~al.}(2015){Bulla}, {Sim}, {Pakmor}, {Kromer},
  {Taubenberger}, {Roepke}, {Hillebrandt}, \& {Seitenzahl}}]{Bulla}
{Bulla}, M., {Sim}, S.~A., {Pakmor}, R., {et~al.} 2015, ArXiv e-prints,
  arXiv:1510.04128

\bibitem[{{Cardelli} {et~al.}(1989){Cardelli}, {Clayton}, \& {Mathis}}]{CCM}
{Cardelli}, J.~A., {Clayton}, G.~C., \& {Mathis}, J.~S. 1989, in IAU Symposium,
  Vol. 135, Interstellar Dust, ed. L.~J. {Allamandola} \& A.~G.~G.~M.
  {Tielens}, 5P

\bibitem[{{Childress} {et~al.}(2013){Childress}, {Scalzo}, {Sim}, {Tucker},
  {Yuan}, {Schmidt}, {Cenko}, {Silverman}, {Contreras}, {Hsiao}, {Phillips},
  {Morrell}, {Jha}, {McCully}, {Filippenko}, {Anderson}, {Benetti}, {Bufano},
  {de Jaeger}, {Forster}, {Gal-Yam}, {Le Guillou}, {Maguire}, {Maund},
  {Mazzali}, {Pignata}, {Smartt}, {Spyromilio}, {Sullivan}, {Taddia},
  {Valenti}, {Bayliss}, {Bessell}, {Blanc}, {Carson}, {Clubb}, {de Burgh-Day},
  {Desjardins}, {Fang}, {Fox}, {Gates}, {Ho}, {Keller}, {Kelly}, {Lidman},
  {Loaring}, {Mould}, {Owers}, {Ozbilgen}, {Pei}, {Pickering}, {Pracy}, {Rich},
  {Schaefer}, {Scott}, {Stritzinger}, {Vogt}, \& {Zhou}}]{Childress}
{Childress}, M.~J., {Scalzo}, R.~A., {Sim}, S.~A., {et~al.} 2013, \apj, 770, 29

\bibitem[{{Chomiuk} {et~al.}(2012){Chomiuk}, {Soderberg}, {Moe}, {Chevalier},
  {Rupen}, {Badenes}, {Margutti}, {Fransson}, {Fong}, \&
  {Dittmann}}]{Chomiuk_2011fe}
{Chomiuk}, L., {Soderberg}, A.~M., {Moe}, M., {et~al.} 2012, \apj, 750, 164

\bibitem[{{Chornock} \& {Filippenko}(2008)}]{SP2004S}
{Chornock}, R., \& {Filippenko}, A.~V. 2008, \aj, 136, 2227

\bibitem[{{Chugai}(1992)}]{Chugai}
{Chugai}, N.~N. 1992, Soviet Astronomy Letters, 18, 168

\bibitem[{{Clayton} {et~al.}(1992){Clayton}, {Anderson}, {Magalhaes}, {Code},
  {Nordsieck}, {Meade}, {Wolff}, {Babler}, {Bjorkman}, {Schulte-Ladbeck},
  {Taylor}, \& {Whitney}}]{UV_1992}
{Clayton}, G.~C., {Anderson}, C.~M., {Magalhaes}, A.~M., {et~al.} 1992, \apjl,
  385, L53

\bibitem[{{Cox} \& {Patat}(2014)}]{SP2008fp}
{Cox}, N.~L.~J., \& {Patat}, F. 2014, \aap, 565, A61

\bibitem[{{Crotts}(2015)}]{CrottsLE}
{Crotts}, A.~P.~S. 2015, \apjl, 804, L37

\bibitem[{{Crotts} \& {Yourdon}(2008)}]{Crotts_LE06X}
{Crotts}, A.~P.~S., \& {Yourdon}, D. 2008, \apj, 689, 1186

\bibitem[{{Dalcanton} {et~al.}(2009){Dalcanton}, {Williams}, {Seth}, {Dolphin},
  {Holtzman}, {Rosema}, {Skillman}, {Cole}, {Girardi}, {Gogarten},
  {Karachentsev}, {Olsen}, {Weisz}, {Christensen}, {Freeman}, {Gilbert},
  {Gallart}, {Harris}, {Hodge}, {de Jong}, {Karachentseva}, {Mateo}, {Stetson},
  {Tavarez}, {Zaritsky}, {Governato}, \& {Quinn}}]{Dalcanton}
{Dalcanton}, J.~J., {Williams}, B.~F., {Seth}, A.~C., {et~al.} 2009, \apjs,
  183, 67

\bibitem[{{Dilday} {et~al.}(2012){Dilday}, {Howell}, {Cenko}, {Silverman},
  {Nugent}, {Sullivan}, {Ben-Ami}, {Bildsten}, {Bolte}, {Endl}, {Filippenko},
  {Gnat}, {Horesh}, {Hsiao}, {Kasliwal}, {Kirkman}, {Maguire}, {Marcy},
  {Moore}, {Pan}, {Parrent}, {Podsiadlowski}, {Quimby}, {Sternberg}, {Suzuki},
  {Tytler}, {Xu}, {Bloom}, {Gal-Yam}, {Hook}, {Kulkarni}, {Law}, {Ofek},
  {Polishook}, \& {Poznanski}}]{Dilday}
{Dilday}, B., {Howell}, D.~A., {Cenko}, S.~B., {et~al.} 2012, Science, 337, 942

\bibitem[{{Elias-Rosa} {et~al.}(2006){Elias-Rosa}, {Benetti}, {Cappellaro},
  {Turatto}, {Mazzali}, {Patat}, {Meikle}, {Stehle}, {Pastorello}, {Pignata},
  {Kotak}, {Harutyunyan}, {Altavilla}, {Navasardyan}, {Qiu}, {Salvo}, \&
  {Hillebrandt}}]{Elias-Rosa_2006}
{Elias-Rosa}, N., {Benetti}, S., {Cappellaro}, E., {et~al.} 2006, \mnras, 369,
  1880

\bibitem[{{Elias-Rosa} {et~al.}(2008){Elias-Rosa}, {Benetti}, {Turatto},
  {Cappellaro}, {Valenti}, {Arkharov}, {Beckman}, {di Paola}, {Dolci},
  {Filippenko}, {Foley}, {Krisciunas}, {Larionov}, {Li}, {Meikle},
  {Pastorello}, {Valentini}, \& {Hillebrandt}}]{Elias-Rosa_2008}
{Elias-Rosa}, N., {Benetti}, S., {Turatto}, M., {et~al.} 2008, \mnras, 384, 107

\bibitem[{{Foley} {et~al.}(2014){Foley}, {Fox}, {McCully}, {Phillips}, {Sand},
  {Zheng}, {Challis}, {Filippenko}, {Folatelli}, {Hillebrandt}, {Hsiao}, {Jha},
  {Kirshner}, {Kromer}, {Marion}, {Nelson}, {Pakmor}, {Pignata}, {R{\"o}pke},
  {Seitenzahl}, {Silverman}, {Skrutskie}, \& {Stritzinger}}]{Foley}
{Foley}, R.~J., {Fox}, O.~D., {McCully}, C., {et~al.} 2014, \mnras, 443, 2887

\bibitem[{{Fossey} {et~al.}(2014){Fossey}, {Cooke}, {Pollack}, {Wilde}, \&
  {Wright}}]{Fossey}
{Fossey}, J., {Cooke}, B., {Pollack}, G., {Wilde}, M., \& {Wright}, T. 2014,
  Central Bureau Electronic Telegrams, 3792, 1

\bibitem[{{Garnavich} {et~al.}(2001){Garnavich}, {Kirshner}, {Challis}, {Jha},
  {Branch}, {Chevalier}, {Filippenko}, {Li}, {Fransson}, {Lundqvist}, {McCray},
  {Panagia}, {Phillips}, {Pun}, {Sonneborn}, {Schmidt}, {Suntzeff}, {Wheeler},
  \& {Supernova INtensive Study SINS Collaboration}}]{Garnavich}
{Garnavich}, P.~M., {Kirshner}, R.~P., {Challis}, P., {et~al.} 2001, in
  Bulletin of the American Astronomical Society, Vol.~33, American Astronomical
  Society Meeting Abstracts, 1370

\bibitem[{{Goobar}(2008)}]{Goobar_Scat}
{Goobar}, A. 2008, \apjl, 686, L103

\bibitem[{{Goobar} {et~al.}(2014){Goobar}, {Johansson}, {Amanullah}, {Cao},
  {Perley}, {Kasliwal}, {Ferretti}, {Nugent}, {Harris}, {Gal-Yam}, {Ofek},
  {Tendulkar}, {Dennefeld}, {Valenti}, {Arcavi}, {Banerjee}, {Venkataraman},
  {Joshi}, {Ashok}, {Cenko}, {Diaz}, {Fremling}, {Horesh}, {Howell},
  {Kulkarni}, {Papadogiannakis}, {Petrushevska}, {Sand}, {Sollerman},
  {Stanishev}, {Bloom}, {Surace}, {Dupuy}, \& {Liu}}]{Goobar}
{Goobar}, A., {Johansson}, J., {Amanullah}, R., {et~al.} 2014, \apjl, 784, L12

\bibitem[{{Greaves} {et~al.}(2000){Greaves}, {Holland}, {Jenness}, \&
  {Hawarden}}]{Greaves}
{Greaves}, J.~S., {Holland}, W.~S., {Jenness}, T., \& {Hawarden}, T.~G. 2000,
  \nat, 404, 732

\bibitem[{{Heiles}(2000)}]{Heiles}
{Heiles}, C. 2000, \aj, 119, 923

\bibitem[{{Hicken} {et~al.}(2009){Hicken}, {Wood-Vasey}, {Blondin}, {Challis},
  {Jha}, {Kelly}, {Rest}, \& {Kirshner}}]{Hicken}
{Hicken}, M., {Wood-Vasey}, W.~M., {Blondin}, S., {et~al.} 2009, \apj, 700,
  1097

\bibitem[{{Hoang}(2015)}]{Hoang}
{Hoang}, T. 2015, ArXiv e-prints, arXiv:1510.01822

\bibitem[{{Hoffman} {et~al.}(2005){Hoffman}, {Chornock}, {Leonard}, \&
  {Filippenko}}]{Hoffman}
{Hoffman}, J.~L., {Chornock}, R., {Leonard}, D.~C., \& {Filippenko}, A.~V.
  2005, \mnras, 363, 1241

\bibitem[{{H\"oflich}(1991)}]{Hoflich}
{H\"oflich}, P. 1991, \aap, 246, 481

\bibitem[{{Hough} {et~al.}(1987){Hough}, {Bailey}, {Rouse}, \&
  {Whittet}}]{SP1986G}
{Hough}, J.~H., {Bailey}, J.~A., {Rouse}, M.~F., \& {Whittet}, D.~C.~B. 1987,
  \mnras, 227, 1P

\bibitem[{{Howell}(2011)}]{Howell_review}
{Howell}, D.~A. 2011, Nature Communications, 2, 350

\bibitem[{{Howell} {et~al.}(2001){Howell}, {H{\"o}flich}, {Wang}, \&
  {Wheeler}}]{SP1999by}
{Howell}, D.~A., {H{\"o}flich}, P., {Wang}, L., \& {Wheeler}, J.~C. 2001, \apj,
  556, 302

\bibitem[{{Hoyle} \& {Fowler}(1960)}]{Hoyle}
{Hoyle}, F., \& {Fowler}, W.~A. 1960, \apj, 132, 565

\bibitem[{{Hutton} {et~al.}(2014){Hutton}, {Ferreras}, {Wu}, {Kuin},
  {Breeveld}, {Yershov}, {Cropper}, \& {Page}}]{Hutton_2014}
{Hutton}, S., {Ferreras}, I., {Wu}, K., {et~al.} 2014, \mnras, 440, 150

\bibitem[{{Hutton} {et~al.}(2015){Hutton}, {Ferreras}, \&
  {Yershov}}]{Hutton_2015}
{Hutton}, S., {Ferreras}, I., \& {Yershov}, V. 2015, \mnras, 452, 1412

\bibitem[{{Iben} \& {Tutukov}(1984)}]{Iben}
{Iben}, Jr., I., \& {Tutukov}, A.~V. 1984, \apjs, 54, 335

\bibitem[{{Johansson} {et~al.}(2014){Johansson}, {Goobar}, {Kasliwal}, {Helou},
  {Masci}, {Tinyanont}, {Jencson}, {Cao}, {Fox}, {Kromer}, {Amanullah},
  {Banerjee}, {Joshi}, {Jerkstrand}, {Kankare}, \& {Prince}}]{Johansson}
{Johansson}, J., {Goobar}, A., {Kasliwal}, M.~M., {et~al.} 2014, ArXiv
  e-prints, arXiv:1411.3332

\bibitem[{{Kasen} {et~al.}(2004){Kasen}, {Nugent}, {Thomas}, \&
  {Wang}}]{Kasen_hole}
{Kasen}, D., {Nugent}, P., {Thomas}, R.~C., \& {Wang}, L. 2004, \apj, 610, 876

\bibitem[{{Kasen} {et~al.}(2003){Kasen}, {Nugent}, {Wang}, {Howell}, {Wheeler},
  {H{\"o}flich}, {Baade}, {Baron}, \& {Hauschildt}}]{Kasen}
{Kasen}, D., {Nugent}, P., {Wang}, L., {et~al.} 2003, \apj, 593, 788

\bibitem[{{Kawabata} {et~al.}(2014){Kawabata}, {Akitaya}, {Yamanaka}, {Itoh},
  {Maeda}, {Moritani}, {Ui}, {Kawabata}, {Mori}, {Nogami}, {Nomoto}, {Suzuki},
  {Takaki}, {Tanaka}, {Ueno}, {Chiyonobu}, {Harao}, {Matsui}, {Miyamoto},
  {Nagae}, {Nakashima}, {Nakaya}, {Ohashi}, {Ohsugi}, {Komatsu}, {Sakimoto},
  {Sasada}, {Sato}, {Tanaka}, {Urano}, {Yamashita}, {Yoshida}, {Arai},
  {Ebisuda}, {Fukazawa}, {Fukui}, {Hashimoto}, {Honda}, {Izumiura}, {Kanda},
  {Kawaguchi}, {Kawai}, {Kuroda}, {Masumoto}, {Matsumoto}, {Nakaoka}, {Takata},
  {Uemura}, \& {Yanagisawa}}]{Kawabata}
{Kawabata}, K.~S., {Akitaya}, H., {Yamanaka}, M., {et~al.} 2014, \apjl, 795, L4

\bibitem[{{Kim} {et~al.}(1994){Kim}, {Martin}, \& {Hendry}}]{Kim}
{Kim}, S.-H., {Martin}, P.~G., \& {Hendry}, P.~D. 1994, \apj, 422, 164

\bibitem[{{Krisciunas} {et~al.}(2006){Krisciunas}, {Prieto}, {Garnavich},
  {Riley}, {Rest}, {Stubbs}, \& {McMillan}}]{Kris_2006}
{Krisciunas}, K., {Prieto}, J.~L., {Garnavich}, P.~M., {et~al.} 2006, \aj, 131,
  1639

\bibitem[{{Leonard} {et~al.}(2001){Leonard}, {Filippenko}, {Ardila}, \&
  {Brotherton}}]{LeonardSP1999em}
{Leonard}, D.~C., {Filippenko}, A.~V., {Ardila}, D.~R., \& {Brotherton}, M.~S.
  2001, \apj, 553, 861

\bibitem[{{Leonard} {et~al.}(2005){Leonard}, {Li}, {Filippenko}, {Foley}, \&
  {Chornock}}]{Leonard}
{Leonard}, D.~C., {Li}, W., {Filippenko}, A.~V., {Foley}, R.~J., \& {Chornock},
  R. 2005, \apj, 632, 450

\bibitem[{{Li} {et~al.}(2011){Li}, {Bloom}, {Podsiadlowski}, {Miller}, {Cenko},
  {Jha}, {Sullivan}, {Howell}, {Nugent}, {Butler}, {Ofek}, {Kasliwal},
  {Richards}, {Stockton}, {Shih}, {Bildsten}, {Shara}, {Bibby}, {Filippenko},
  {Ganeshalingam}, {Silverman}, {Kulkarni}, {Law}, {Poznanski}, {Quimby},
  {McCully}, {Patel}, {Maguire}, \& {Shen}}]{Li_2011feProg}
{Li}, W., {Bloom}, J.~S., {Podsiadlowski}, P., {et~al.} 2011, \nat, 480, 348

\bibitem[{{Livio} \& {Pringle}(2011)}]{Livio}
{Livio}, M., \& {Pringle}, J.~E. 2011, \apjl, 740, L18

\bibitem[{{Maeda} {et~al.}(2016){Maeda}, {Tajitsu}, {Kawabata}, {Foley},
  {Honda}, {Moritani}, {Tanaka}, {Hashimoto}, {Ishigaki}, {Simon}, {Phillips},
  {Yamanaka}, {Nogami}, {Arai}, {Aoki}, {Nomoto}, {Milisavljevic}, {Mazzali},
  {Soderberg}, {Schramm}, {Sato}, {Harakawa}, {Morrell}, \&
  {Arimoto}}]{Maeda_ISM}
{Maeda}, K., {Tajitsu}, A., {Kawabata}, K.~S., {et~al.} 2016, \apj, 816, 57

\bibitem[{{Maoz} {et~al.}(2014){Maoz}, {Mannucci}, \& {Nelemans}}]{Maoz_ARAA}
{Maoz}, D., {Mannucci}, F., \& {Nelemans}, G. 2014, \araa, 52, 107

\bibitem[{{Margutti} {et~al.}(2012){Margutti}, {Soderberg}, {Chomiuk},
  {Chevalier}, {Hurley}, {Milisavljevic}, {Foley}, {Hughes}, {Slane},
  {Fransson}, {Moe}, {Barthelmy}, {Boynton}, {Briggs}, {Connaughton}, {Costa},
  {Cummings}, {Del Monte}, {Enos}, {Fellows}, {Feroci}, {Fukazawa}, {Gehrels},
  {Goldsten}, {Golovin}, {Hanabata}, {Harshman}, {Krimm}, {Litvak},
  {Makishima}, {Marisaldi}, {Mitrofanov}, {Murakami}, {Ohno}, {Palmer},
  {Sanin}, {Starr}, {Svinkin}, {Takahashi}, {Tashiro}, {Terada}, \&
  {Yamaoka}}]{Margutti_2011fe}
{Margutti}, R., {Soderberg}, A.~M., {Chomiuk}, L., {et~al.} 2012, \apj, 751,
  134

\bibitem[{{Marietta} {et~al.}(2000){Marietta}, {Burrows}, \&
  {Fryxell}}]{Marietta}
{Marietta}, E., {Burrows}, A., \& {Fryxell}, B. 2000, \apjs, 128, 615

\bibitem[{{Marion} {et~al.}(2015){Marion}, {Sand}, {Hsiao}, {Banerjee},
  {Valenti}, {Stritzinger}, {Vink{\'o}}, {Joshi}, {Venkataraman}, {Ashok},
  {Amanullah}, {Binzel}, {Bochanski}, {Bryngelson}, {Burns}, {Drozdov},
  {Fieber-Beyer}, {Graham}, {Howell}, {Johansson}, {Kirshner}, {Milne},
  {Parrent}, {Silverman}, {Vervack}, \& {Wheeler}}]{Marion}
{Marion}, G.~H., {Sand}, D.~J., {Hsiao}, E.~Y., {et~al.} 2015, \apj, 798, 39

\bibitem[{{Martin} {et~al.}(1999){Martin}, {Clayton}, \& {Wolff}}]{UV_1999}
{Martin}, P.~G., {Clayton}, G.~C., \& {Wolff}, M.~J. 1999, \apj, 510, 905

\bibitem[{{Maund} {et~al.}(2010){Maund}, {H{\"o}flich}, {Patat}, {Wheeler},
  {Zelaya}, {Baade}, {Wang}, {Clocchiatti}, \& {Quinn}}]{Maund_uni}
{Maund}, J.~R., {H{\"o}flich}, P., {Patat}, F., {et~al.} 2010, \apjl, 725, L167

\bibitem[{{Maund} {et~al.}(2013){Maund}, {Spyromilio}, {H{\"o}flich},
  {Wheeler}, {Baade}, {Clocchiatti}, {Patat}, {Reilly}, {Wang}, \&
  {Zelaya}}]{SP2012fr}
{Maund}, J.~R., {Spyromilio}, J., {H{\"o}flich}, P.~A., {et~al.} 2013, \mnras,
  433, L20

\bibitem[{{McCall}(1984)}]{McCall}
{McCall}, M.~L. 1984, \mnras, 210, 829

\bibitem[{{Milne} {et~al.}(2016){Milne}, {Williams}, {Smith}, {Porter},
  {Smith}, D., {Jannuzi}, \& {Green}}]{SP2011fe}
{Milne}, P., {Williams}, G.~G., {Smith}, P.~S., {et~al.} 2016, private
  communication

\bibitem[{{Patat} {et~al.}(2009){Patat}, {Baade}, {H{\"o}flich}, {Maund},
  {Wang}, \& {Wheeler}}]{SP2006X}
{Patat}, F., {Baade}, D., {H{\"o}flich}, P., {et~al.} 2009, Astronomy and
  Astrophysics, 508, 229

\bibitem[{{Patat} {et~al.}(2012){Patat}, {H{\"o}flich}, {Baade}, {Maund},
  {Wang}, \& {Wheeler}}]{SP2005ke}
{Patat}, F., {H{\"o}flich}, P., {Baade}, D., {et~al.} 2012, \aap, 545, A7

\bibitem[{{Patat} {et~al.}(2007){Patat}, {Chandra}, {Chevalier}, {Justham},
  {Podsiadlowski}, {Wolf}, {Gal-Yam}, {Pasquini}, {Crawford}, {Mazzali},
  {Pauldrach}, {Nomoto}, {Benetti}, {Cappellaro}, {Elias-Rosa}, {Hillebrandt},
  {Leonard}, {Pastorello}, {Renzini}, {Sabbadin}, {Simon}, \&
  {Turatto}}]{Patat_CSM}
{Patat}, F., {Chandra}, P., {Chevalier}, R., {et~al.} 2007, Science, 317, 924

\bibitem[{{Patat} {et~al.}(2014){Patat}, {Taubenberger}, {Baade}, {Hoeflich},
  {Maund}, {Reilly}, {Spyromilio}, {Wang}, {Wheeler}, \& {Zelaya}}]{Patat_ATEL}
{Patat}, F., {Taubenberger}, S., {Baade}, D., {et~al.} 2014, The Astronomer's
  Telegram, 5830, 1

\bibitem[{{Patat} {et~al.}(2015){Patat}, {Taubenberger}, {Cox}, {Baade},
  {Clocchiatti}, {H{\"o}flich}, {Maund}, {Reilly}, {Spyromilio}, {Wang},
  {Wheeler}, \& {Zelaya}}]{Patat_Dust}
{Patat}, F., {Taubenberger}, S., {Cox}, N.~L.~J., {et~al.} 2015, \aap, 577, A53

\bibitem[{{Perlmutter} {et~al.}(1999){Perlmutter}, {Aldering}, {Goldhaber},
  {Knop}, {Nugent}, {Castro}, {Deustua}, {Fabbro}, {Goobar}, {Groom}, {Hook},
  {Kim}, {Kim}, {Lee}, {Nunes}, {Pain}, {Pennypacker}, {Quimby}, {Lidman},
  {Ellis}, {Irwin}, {McMahon}, {Ruiz-Lapuente}, {Walton}, {Schaefer}, {Boyle},
  {Filippenko}, {Matheson}, {Fruchter}, {Panagia}, {Newberg}, {Couch}, \&
  {Project}}]{Perlmutter}
{Perlmutter}, S., {Aldering}, G., {Goldhaber}, G., {et~al.} 1999, \apj, 517,
  565

\bibitem[{{Phillips} {et~al.}(1999){Phillips}, {Lira}, {Suntzeff}, {Schommer},
  {Hamuy}, \& {Maza}}]{Phillips_Color}
{Phillips}, M.~M., {Lira}, P., {Suntzeff}, N.~B., {et~al.} 1999, \aj, 118, 1766

\bibitem[{{Phillips} {et~al.}(2013){Phillips}, {Simon}, {Morrell}, {Burns},
  {Cox}, {Foley}, {Karakas}, {Patat}, {Sternberg}, {Williams}, {Gal-Yam},
  {Hsiao}, {Leonard}, {Persson}, {Stritzinger}, {Thompson}, {Campillay},
  {Contreras}, {Folatelli}, {Freedman}, {Hamuy}, {Roth}, {Shields}, {Suntzeff},
  {Chomiuk}, {Ivans}, {Madore}, {Penprase}, {Perley}, {Pignata}, {Preston}, \&
  {Soderberg}}]{Phillips_Na}
{Phillips}, M.~M., {Simon}, J.~D., {Morrell}, N., {et~al.} 2013, \apj, 779, 38

\bibitem[{{Riess} {et~al.}(1998){Riess}, {Filippenko}, {Challis},
  {Clocchiatti}, {Diercks}, {Garnavich}, {Gilliland}, {Hogan}, {Jha},
  {Kirshner}, {Leibundgut}, {Phillips}, {Reiss}, {Schmidt}, {Schommer},
  {Smith}, {Spyromilio}, {Stubbs}, {Suntzeff}, \& {Tonry}}]{Riess}
{Riess}, A.~G., {Filippenko}, A.~V., {Challis}, P., {et~al.} 1998, \aj, 116,
  1009

\bibitem[{{Schaefer} \& {Pagnotta}(2012)}]{Schaefer}
{Schaefer}, B.~E., \& {Pagnotta}, A. 2012, \nat, 481, 164

\bibitem[{{Schmidt} {et~al.}(1992{\natexlab{a}}){Schmidt}, {Elston}, \&
  {Lupie}}]{Schmidt}
{Schmidt}, G.~D., {Elston}, R., \& {Lupie}, O.~L. 1992{\natexlab{a}}, \aj, 104,
  1563

\bibitem[{{Schmidt} {et~al.}(1992{\natexlab{b}}){Schmidt}, {Stockman}, \&
  {Smith}}]{SPOL}
{Schmidt}, G.~D., {Stockman}, H.~S., \& {Smith}, P.~S. 1992{\natexlab{b}},
  \apjl, 398, L57

\bibitem[{{Serkowski} {et~al.}(1975){Serkowski}, {Mathewson}, \&
  {Ford}}]{Serkowski1975}
{Serkowski}, K., {Mathewson}, D.~S., \& {Ford}, V.~L. 1975, \apj, 196, 261

\bibitem[{{Silverman} {et~al.}(2015){Silverman}, {Vink{\'o}}, {Marion},
  {Wheeler}, {Barna}, {Szalai}, {Mulligan}, \& {Filippenko}}]{Silverman}
{Silverman}, J.~M., {Vink{\'o}}, J., {Marion}, G.~H., {et~al.} 2015, \mnras,
  451, 1973

\bibitem[{{Simon} {et~al.}(2009){Simon}, {Gal-Yam}, {Gnat}, {Quimby},
  {Ganeshalingam}, {Silverman}, {Blondin}, {Li}, {Filippenko}, {Wheeler},
  {Kirshner}, {Patat}, {Nugent}, {Foley}, {Vogt}, {Butler}, {Peek},
  {Rosolowsky}, {Herczeg}, {Sauer}, \& {Mazzali}}]{Simon}
{Simon}, J.~D., {Gal-Yam}, A., {Gnat}, O., {et~al.} 2009, \apj, 702, 1157

\bibitem[{{Stehle} {et~al.}(2005){Stehle}, {Mazzali}, {Benetti}, \&
  {Hillebrandt}}]{Stehle_2002bo}
{Stehle}, M., {Mazzali}, P.~A., {Benetti}, S., \& {Hillebrandt}, W. 2005,
  \mnras, 360, 1231

\bibitem[{{Sternberg} {et~al.}(2014){Sternberg}, {Gal-Yam}, {Simon}, {Patat},
  {Hillebrandt}, {Phillips}, {Foley}, {Thompson}, {Morrell}, {Chomiuk},
  {Soderberg}, {Yong}, {Kraus}, {Herczeg}, {Hsiao}, {Raskutti}, {Cohen},
  {Mazzali}, \& {Nomoto}}]{Sternberg_2014}
{Sternberg}, A., {Gal-Yam}, A., {Simon}, J.~D., {et~al.} 2014, \mnras, 443,
  1849

\bibitem[{{Tanaka} {et~al.}(2010){Tanaka}, {Kawabata}, {Yamanaka}, {Maeda},
  {Hattori}, {Aoki}, {Nomoto}, {Iye}, {Sasaki}, {Mazzali}, \&
  {Pian}}]{SP2009dc}
{Tanaka}, M., {Kawabata}, K.~S., {Yamanaka}, M., {et~al.} 2010, \apj, 714, 1209

\bibitem[{{Trammell} {et~al.}(1993){Trammell}, {Dinerstein}, \&
  {Goodrich}}]{TrammellRSP}
{Trammell}, S.~R., {Dinerstein}, H.~L., \& {Goodrich}, R.~W. 1993, \apj, 402,
  249

\bibitem[{{Vallely} {et~al.}(2015){Vallely}, {Moreno-Raya}, {Baron},
  {Ruiz-Lapuente}, {Dominguez}, {Galbany}, {Gonzalez Hernandez}, {Mendez},
  {Hamuy}, {Lopez-Sanchez}, {Catalan}, {Cooke}, {Farina}, {Genova-Santos},
  {Karjalainen}, {Lietzen}, {McCormac}, {Riddick}, {Rubino-Martin}, {Tudor}, \&
  {Vaduvescu}}]{Vallely}
{Vallely}, P., {Moreno-Raya}, M.~E., {Baron}, E., {et~al.} 2015, ArXiv
  e-prints, arXiv:1512.02608

\bibitem[{{Voshchinnikov} {et~al.}(2012){Voshchinnikov}, {Henning},
  {Prokopjeva}, \& {Das}}]{Voshchinnikov}
{Voshchinnikov}, N.~V., {Henning}, T., {Prokopjeva}, M.~S., \& {Das}, H.~K.
  2012, \aap, 541, A52

\bibitem[{{Wang}(2005)}]{WangLE}
{Wang}, L. 2005, \apjl, 635, L33

\bibitem[{{Wang} {et~al.}(2007){Wang}, {Baade}, \& {Patat}}]{Wang_Sci}
{Wang}, L., {Baade}, D., \& {Patat}, F. 2007, Science, 315, 212

\bibitem[{{Wang} \& {Wheeler}(2008)}]{SPReview}
{Wang}, L., \& {Wheeler}, J.~C. 2008, \araa, 46, 433

\bibitem[{{Wang} {et~al.}(1997){Wang}, {Wheeler}, \& {H{\"o}flich}}]{SP1996X}
{Wang}, L., {Wheeler}, J.~C., \& {H{\"o}flich}, P. 1997, \apjl, 476, L27

\bibitem[{{Wang} {et~al.}(2003){Wang}, {Baade}, {H{\"o}flich}, {Khokhlov},
  {Wheeler}, {Kasen}, {Nugent}, {Perlmutter}, {Fransson}, \&
  {Lundqvist}}]{SP2001el}
{Wang}, L., {Baade}, D., {H{\"o}flich}, P., {et~al.} 2003, \apj, 591, 1110

\bibitem[{{Wang} {et~al.}(2008{\natexlab{a}}){Wang}, {Li}, {Filippenko},
  {Foley}, {Smith}, \& {Wang}}]{Wang_LE06X}
{Wang}, X., {Li}, W., {Filippenko}, A.~V., {et~al.} 2008{\natexlab{a}}, \apj,
  677, 1060

\bibitem[{{Wang} {et~al.}(2008{\natexlab{b}}){Wang}, {Li}, {Filippenko},
  {Krisciunas}, {Suntzeff}, {Li}, {Zhang}, {Deng}, {Foley}, {Ganeshalingam},
  {Li}, {Lou}, {Qiu}, {Shang}, {Silverman}, {Zhang}, \& {Zhang}}]{XWang_2006X}
---. 2008{\natexlab{b}}, \apj, 675, 626

\bibitem[{{Wang} {et~al.}(2009){Wang}, {Li}, {Filippenko}, {Foley}, {Kirshner},
  {Modjaz}, {Bloom}, {Brown}, {Carter}, {Friedman}, {Gal-Yam}, {Ganeshalingam},
  {Hicken}, {Krisciunas}, {Milne}, {Silverman}, {Suntzeff}, {Wood-Vasey},
  {Cenko}, {Challis}, {Fox}, {Kirkman}, {Li}, {Li}, {Malkan}, {Moore},
  {Reitzel}, {Rich}, {Serduke}, {Shang}, {Steele}, {Swift}, {Tao}, {Wong}, \&
  {Zhang}}]{Wang_golden}
---. 2009, \apj, 697, 380

\bibitem[{{Wardle} \& {Kronberg}(1974)}]{Wardle}
{Wardle}, J.~F.~C., \& {Kronberg}, P.~P. 1974, \apj, 194, 249

\bibitem[{{Webbink}(1984)}]{Webbink}
{Webbink}, R.~F. 1984, \apj, 277, 355

\bibitem[{{Welty} {et~al.}(2014){Welty}, {Ritchey}, {Dahlstrom}, \&
  {York}}]{Welty}
{Welty}, D.~E., {Ritchey}, A.~M., {Dahlstrom}, J.~A., \& {York}, D.~G. 2014,
  \apj, 792, 106

\bibitem[{{Whelan} \& {Iben}(1973)}]{Whelan}
{Whelan}, J., \& {Iben}, Jr., I. 1973, \apj, 186, 1007

\bibitem[{Whittet(1992)}]{Whittet_book}
Whittet, D. C.~B. 1992, Dust in the Galactic Environment (Institute of Physics
  Publishing)

\bibitem[{{Whittet} {et~al.}(1992){Whittet}, {Martin}, {Hough}, {Rouse},
  {Bailey}, \& {Axon}}]{Whittet1992}
{Whittet}, D.~C.~B., {Martin}, P.~G., {Hough}, J.~H., {et~al.} 1992, \apj, 386,
  562

\bibitem[{{Yang} {et~al.}(2015){Yang}, {Wang}, {Baade}, {Brown}, {Clocchiatti},
  {Cracraft}, {Hoflich}, {Maund}, {Patat}, {Sparks}, {Spyromilio}, {Wang}, \&
  {Wheeler}}]{Yang_Echo}
{Yang}, Y., {Wang}, L., {Baade}, D., {et~al.} 2015, ArXiv e-prints,
  arXiv:1511.02495

\bibitem[{{Zelaya} {et~al.}(2013){Zelaya}, {Quinn}, {Baade}, {Clocchiatti},
  {H{\"o}flich}, {Maund}, {Patat}, {Wang}, \& {Wheeler}}]{SP2007sr}
{Zelaya}, P., {Quinn}, J.~R., {Baade}, D., {et~al.} 2013, \aj, 145, 27

\end{thebibliography}
\clearpage

\end{document}